\documentclass[twocolumn]{aa}

\usepackage{graphicx}
\usepackage{txfonts}
\usepackage{epsfig}
\usepackage{amsbsy}
\usepackage{amssymb}
\usepackage{amsmath}
\usepackage{latexsym}
\usepackage{natbib}
\usepackage{verbatim}
\usepackage{bm}
\usepackage{tabularx}
\usepackage{soul}
\usepackage{xcolor}
\usepackage{hyperref}
\usepackage{placeins}
\usepackage{soul}
\usepackage{subfig}

\hypersetup{
  colorlinks   = true, %Colours links instead of ugly boxes
  urlcolor     = blue, %Colour for external hyperlinks
  linkcolor    = blue, %Colour of internal links
  citecolor    = blue  %Colour of citations
}

\begin{document}

% \title{3D modeling of scattering polarization in the Ca~{\sc i} 4227\,{\AA} line\\ 
% with angle-dependent partial frequency redistribution}
% \subtitle{I. Verification and forward scattering}

\title{3D radiative transfer modeling of scattering polarization \\ with partial frequency redistribution}
\subtitle{I. Verification and disk-center results for the solar Ca~{\sc i} 4227\,{\AA} line}

\titlerunning{3D modeling of scattering polarization with PRD}
%Scattering polarization in Ca~{\sc i} 4227 with AD PRD
\authorrunning{Benedusi et al.}

\author{
    Pietro Benedusi\inst{1,2}
	\and
    Simone Riva\inst{2,1}
	\and
    Tanaus\'u del Pino Alem\'an\inst{3,4}
	\and
    Gioele Janett\inst{1,2}
    \and \\
    Fabio Riva\inst{1,2}
	\and
    Ji\v{r}\'i \v{S}t\v{e}p\'an\inst{5}
    \and
    Rolf Krause\inst{6,7,2}
    \and
    Javier Trujillo Bueno\inst{3,4,8}
    \and
    Luca Belluzzi\inst{1,2}
}

\institute{
    Istituto ricerche solari Aldo e Cele Dacc\`o (IRSOL), Faculty of Informatics, Universit\`a della Svizzera italiana, Locarno, Switzerland
    \and
    Euler Institute, Faculty of Informatics, Universit\`a della Svizzera italiana, Lugano, Switzerland
    \and
    Instituto de Astrof\'isica de Canarias, La Laguna, Tenerife, Spain
	\and
	Departamento de Astrof\'isica, Universidad de La Laguna, La Laguna, Tenerife, Spain
    \and
    Astronomical Institute ASCR, Ond\v{r}ejov, Czech Republic
    \and
    AMCS, KAUST, Kinga Abdullah University of Science and Technology, Thuwal, Saudi-Arabia
	\and
    FernUni, Brig, Switzerland
    \and
	Consejo Superior de Investigaciones Cient\'ificas, Spain\\
	\email{pietro.benedusi@usi.ch}
}

\newcommand{\corange}[1]{\textcolor{orange}{#1}}
\newcommand{\cred}[1]{\textcolor{red}{#1}}
\newcommand{\MyNewEdit}[1]{\textcolor{green}{#1}}
\newcommand{\eab}[1]{\textcolor{teal}{#1}}
\newcommand{\cluca}[1]{\textcolor{teal}{#1}}
\newcommand{\cgreen}[1]{\textcolor{green}{#1}}

\abstract
{
Several strong solar resonance lines show observable linear scattering polarization signals, holding a great potential for investigating the magnetism of the outer solar atmosphere.
Accurately modeling these signals requires solving the radiative transfer (RT) problem for polarized radiation in comprehensive 3D models of the solar atmosphere, in non-local thermodynamic equilibrium (non-LTE), accounting for partial frequency redistribution (PRD) effects.
This problem has so far been computationally inaccessible.
}
{
We present the first scientific application of TRIP, a novel software for the massively parallel solution of the 3D non-LTE RT problem for polarized radiation, including scattering polarization and PRD.
We aim to verify the code and explore the combined action of PRD and the 3D structure of the solar atmosphere on scattering polarization.
}
{
We run TRIP to synthesize the Stokes profiles of the Ca~{\sc i} line at 4227\,{\AA} in a 3D model of the solar atmosphere extracted from a radiation magneto-hydrodynamic simulation.
%\cred{\st{We consider a two-level atomic model and we assume that the population of the lower level is a fixed input parameter.}}
We efficiently solve the resulting large-scale problem, with up to $4 \times 10^{10}$ degrees of freedom, with a state-of-the-art preconditioned Krylov method, using up to 20 thousand parallel CPUs.
}
{
% By considering the limit of complete frequency redistribution (CRD), we successfully verify TRIP against \cred{\st{the PORTA RT code} PORTA [remove?]}. 
%
%As a verification step,
We successfully cross-checked TRIP with the established PORTA code, considering the limit of complete frequency redistribution (CRD). We then analyzed how the solution tolerance and the angular discretization affect the PRD results.
%We successfully verified TRIP cross-checking its output in the limit of complete frequency redistribution (CRD) with the well-established PORTA code.
%Subsequently, we verified the accuracy of the PRD results for the chosen tolerances and angular discretization in the PRD setting. 
%
%confirming the accuracy of the TRIP results and providing a cross-verification of both codes. 
%Focusing on the radiation emergent along the vertical, %corresponding to a disk-center observation.
From a physical perspective,
we found that the joint impact of PRD effects and the detailed 3D structure of the atmospheric model produce disk-center scattering polarization signals in the line wings. These signals are sensitive to the magnetic field, via magneto-optical effects, and to bulk velocity gradients.
We also show that %the CRD assumption underestimates 
the CRD approximation underestimates the amplitude of disk-center line-core signals.
}
{
We demonstrate the capabilities of TRIP to accurately model the polarization of strong solar resonance lines in comprehensive 3D atmospheric models. 
This achievement represents a crucial step forward for diagnosing the magnetism of the solar chromosphere and transition region through the quantitative comparisons of synthetic and observational data. 
Ongoing development of TRIP will allow us to consider larger atmospheric models as well as a wider range of chromospheric lines of diagnostic interest. 
%such as Mg~{\sc ii} h \& k, H~{\sc i} Ly-$\alpha$, and He~{\sc ii} at 304\,{\AA}.
}


\maketitle

\section{Introduction}
Linear polarization signals resulting from the scattering of anisotropic radiation (a.k.a. scattering polarization) can be measured in many lines across the solar spectrum, from the near infrared to the far ultraviolet \citep[e.g.,][]{gandorfer2000,gandorfer2002,gandorfer2005,kano2017,rachmeler2022}. 
These signals contain valuable information about the physical conditions of the solar atmosphere, particularly regarding its magnetic fields \citep[e.g.,][]{jtb2001hanle,jtb2022ar}.

The amplitude of scattering polarization signals is primarily governed by the degree of anisotropy of the radiation illuminating the atoms, which is sensitive to the intricate 3D structure of the solar atmosphere and to the presence of plasma bulk velocities \citep[e.g.,][]{manso2011,stepan2016}.
Accurately modeling the scattering polarization in spectral lines thus necessitates solving the radiative transfer (RT) problem for polarized radiation under non-local thermodynamic equilibrium (non-LTE) conditions within comprehensive 3D models of the solar atmosphere.
The development of the PORTA (POlarized Radiative TrAnsfer) code \citep{stepan2013} has revolutionized the numerical modeling of %scattering polarization 
these signals, enabling the solution of the 3D non-LTE RT problem for polarized radiation, %including 
treating scattering processes in the limit of complete frequency redistribution (CRD). 
PORTA can accurately model the polarization of photospheric lines \citep[e.g.,][]{delpino2018}, subordinate chromospheric lines \citep[e.g.,][]{stepan2016}, and it also provides approximate results for the line core polarization signals of strong resonance lines, such as H~{\sc i} Ly-$\alpha$ \citep{stepan2015} and Ca~{\sc i} 4227 \citep{jaume2021CaI}.

To accurately model the polarization of strong resonance lines, it is however essential to include partial frequency redistribution (PRD) effects in the scattering processes.
These effects are responsible for the conspicuous scattering polarization signals that strong resonance lines typically exhibit in their extended wings \citep[e.g.,][]{faurobert92,belluzzi2012a}, and also affect the line-core amplitudes \citep[e.g.,][]{sampoorna2017,belluzzi2024}.
Accounting for PRD effects is notoriously demanding from the computational point of view, and significant efforts have been focused on developing efficient methods to model scattering polarization including  such effects \citep[e.g.,][]{belluzzi2014,nagendra2019,delpino2020,riva2023,riva2025arxiv}.
%Recent studies, mainly carried out in 1D plane-parallel atmospheric models,
Various studies have highlighted the importance of considering PRD effects in their most general angle-dependent (AD) formulation, as the simplifying angle-average (AA) approximation \citep[e.g.,][]{rees1982} can lead to significant inaccuracies \citep[e.g.,][]{sampoorna2017,janett2021a,delpino2025}.
The just quoted works were all carried out in 1D plane-parallel atmospheric models. 
As a matter of fact, only a handful of investigations have so far explored scattering polarization with PRD effects in multi-D geometries \citep[e.g.,][and preceding papers of the series]{anusha2012}, always considering academic atmospheric models to lower the computational complexity of the problem.
In a more recent work, \citet{anusha2023} first modeled scattering polarization with PRD in a static and homogeneous self-emitting 3D model, finding that the AA approximation provides rather accurate results in the unmagnetized case, while it can introduce significant errors when a magnetic field is present.
%By considering a homogeneous 3D box of finite extension, \citet{anusha2023} showed that the AA approximation can introduce significant inaccuracies in the scattering polarization signals, especially in the presence of a magnetic field}.

Formulating the problem within the framework of the PRD theory of \citet{bommier1997a,bommier1997b},
% \footnote{This theory considers a two-level atom with an unpolarized and infinitely sharp lower level.}
\citet{benedusi2023} developed an innovative approach for solving the so far computationally inaccessible %3D 
non-LTE RT problem for polarized radiation, with scattering polarization and AD PRD effects, in comprehensive 3D models of the solar atmosphere.
% The solution strategy assumes a fixed population of the lower level of the line transition, thus rendering the non-LTE RT problem linear w.r.t. the radiation field \citep[e.g.,][]{janett2021b}.
The solution strategy, exploiting an efficient parallel matrix-free Krylov solver with a physics-based preconditioner \citep[see][]{benedusi2021,benedusi2022,janett2024}, is implemented in the TRIP (Three-dimensional Radiative transfer Including Polarization and PRD) code.
In this paper, we present TRIP and we first apply it to model the intensity and polarization of the Ca~{\sc i} line at 4227\,{\AA} in a 3D model of the solar atmosphere extracted from a radiation magneto-hydrodynamic (R-MHD) simulation.
% These calculations allow us to explore, for the first time, the sensitivity of the broad wing scattering polarization signals of this line to the joint action of the 3D structure of the solar atmosphere, magnetic fields, and macroscopic plasma bulk velocities \cluca{[!]}.
In Section~\ref{sec:model}, we outline the non-LTE RT problem for polarized radiation, %\cluca{along with its discretization and algebraic formulation, 
and the applied solution strategy.
%the considered atmospheric model, the problem discretization and algebraic formulation, and the computational approach.
Section~\ref{sec:verification} presents some verification tests of TRIP. 
In Section~\ref{sec:results}, we show the numerical results, focusing on the radiation emergent along the vertical. 
Finally, Section~\ref{sec:conclusions} contains some concluding remarks.

\section{Problem formulation and solution strategy}\label{sec:model}
In this section, we first present the non-LTE RT problem for polarized radiation, along with its discretization and algebraic formulation.
%and describe the considered 3D atmospheric model. 
We then recall the adopted solution strategy, present the considered atmospheric and atomic models, and comment on the employed computational resources.
%We then provide a brief overview of its discretization, algebraic formulation, initialization, and the adopted solution strategy.}

\subsection{Non-LTE RT problem for polarized radiation}\label{sec:formulation}
%A \cred{partially-polarized} radiation field %can be 
%\cred{is usually} described by the four Stokes parameters, \cred{which can be} 
We describe the radiation field through the four Stokes parameters, which can be arranged in the Stokes vector $\bm{I} = (I, Q, U, V)^T$.
Under stationary conditions, assumed throughout this work, the propagation in a given domain $D\subset \mathbb{R}^3$ of a radiation beam of frequency $\nu \in \mathbb{R}_+$ and direction corresponding to the unit vector $\bm{\Omega}=(\theta,\chi)\in[0,\pi]\times[0,2\pi)$, at the spatial point $\bm{r} \in D $, is described by the following RT equation
\begin{equation}\label{eq:RT}
        \vec{\nabla}_{{\vec{\Omega}}}\bm{I}(\bm{r},\vec{\Omega},\nu) = 
        - K(\bm{r},\vec{\Omega},\nu) \bm{I}(\bm{r},\vec{\Omega},\nu) + 
        \pmb{\varepsilon}(\bm{r},\vec{\Omega},\nu),
\end{equation}
where $\vec{\nabla}_{\vec{\Omega}}$ denotes the directional derivative along $\vec{\Omega}$, $K \in\mathbb{R}^{4\times4}$ is the propagation matrix, and $\pmb{\varepsilon}\in\mathbb{R}^{4}$ is the emission vector \citep[e.g.,][]{LL04}.
Both $\pmb{\varepsilon}$ and $K$ depend on the state of the atoms giving rise to the spectral line of interest, which in turn depends on the incident radiation field through the statistical equilibrium (SE) equations.
The considered non-LTE RT problem consists in finding a self-consistent solution of Eq.~\eqref{eq:RT} for the polarized radiation field and of the SE equations for the polarized atomic system.

In this work, we aim to model the Stokes profiles of the Ca~{\sc i} resonance line at 4227\,{\AA}.
The polarization of this line can be suitably synthesized considering an atomic model of
% neutral calcium composed of two levels,
% and assuming that the lower level, that is the ground level of Ca~{\sc i}, is unpolarized and infinitely-sharp
neutral calcium composed of two levels, with an infinitely-sharp and unpolarized lower level
\citep[e.g.,][]{alsina2018CaI,belluzzi2024}.
%\cred{The atomic data are reported in Table~1 \cred{[TO BE PREPARED]}.}
If stimulated emission is neglected, the SE equations for this atomic model can be solved analytically and the emission vector directly relates
to the radiation field illuminating the atoms through:
\begin{align}\label{eq:emission}
        \pmb{\varepsilon}(\bm{r},\vec{\Omega},\nu) = & \, 
        k_L(\bm{r})\!\! \int\!\! {\rm d} \nu'\! 
        \oint\! \frac{{\rm d} \vec{\Omega}'}{4 \pi}
        R(\bm{r},\vec{\Omega}',\vec{\Omega},\nu',\nu) 
        \bm{I}(\bm{r},\vec{\Omega}',\nu') \nonumber \\
        & + \pmb{\varepsilon}^{\text{th}}(\bm{r},\vec{\Omega},\nu),
\end{align}
where $k_L$ is the frequency-integrated absorption coefficient, $R$ is the two-level atom redistribution matrix as derived by \citet{bommier1997a,bommier1997b}, and $\pmb{\varepsilon}^{\text{th}}$ describes the contribution from atoms that are collisionally excited (thermal term).
The integral in \eqref{eq:emission} is known as the scattering integral.
The explicit expressions of $K$ and $\pmb{\varepsilon}^{\text{th}}$ for the considered atomic model can be found,
for instance,
% in \citet{alsina2017} \cluca{[!]} as well as 
in \citet{riva2023}.

In this work, we assume that the population of the lower level is a fixed input parameter of the problem.
This assumption allows us to calculate $K$, $k_L$, and $\pmb{\varepsilon}^{\text{th}}$ a-priori, which in turn renders the whole non-LTE RT problem linear w.r.t. $\bm{I}$ \citep[e.g.,][]{janett2021b}.
This approach has been validated by \citet{riva2025arxiv} in 1D.

\subsection{Discretization}\label{sec:discretization}
%\cluca{One of the main difficulties in solving the non-LTE RT problem including AD PRD effects is the need to keep in memory the full radiation field for the considered discretization of the continuous variables $(\bm{r},\vec{\Omega},\nu)$.}
We formulate the problem in a right-handed Cartesian reference system with the $z$ axis directed along the vertical. %and the $x$ axis directed as shown in Fig.~\ref{fig:atmos_mirror}
The structured discretization of the spatial coordinate $\bm{r}=(x,y,z)$ is provided by the considered atmospheric model.
In state-of-the-art models of the solar atmosphere with approximately $500^3$ spatial nodes (and for typical angular and spectral discretizations) the complete polarized radiation field can easily exceed $10^{12}$ degrees of freedom. 
Notably, when AD PRD effects are taken into account, this is also the number of problem unknowns.
Solving linear systems of such size is currently unfeasible, since $10^{11}$ degrees of freedom is the theoretical limit territory of current HPC platforms.
To keep the problem within this limit, we consider a relatively small 3D atmospheric model, with a spatial grid spanning
$[x_{\min},x_{\max}]= [y_{\min},y_{\max}] = [0\,\mathrm{Mm},5.905\,\mathrm{Mm}]$ and $[z_{\min},z_{\max}] = [-0.114\,\mathrm{Mm},2.520\,\mathrm{Mm}]$ with $\{N_x, N_y, N_z\} = \{63,63,134\}$ grid points. 
More details about the atmospheric model and the way we obtained it are provided in Sect.~\ref{sec:atmosphere}.
We consider the spectral interval $[\lambda_{\min},\lambda_{\max}]=[4221.6\,{\AA}, 4232.7\,{\AA}]$, where wavelengths are in air, discretized with $N_\lambda=96$ unevenly spaced points (with higher resolution in the line core).
%\cred{\st{For the propagation directions $\bm{\Omega} = (\theta,\chi) \in [0,\pi] \times [0,2\pi)$, with $\theta$ the inclination and $\chi$ the azimuth, we consider an angular discretization with $4 \times 4$ directions per octant.
%This is given by the tensor product of 16 directions equally spaced for $\chi$ (i.e., $\chi_i=(i-0.5)2\pi/16$ for $i=1,\ldots,16$), and 8 Gauss-Legendre quadrature nodes in the interval $(-1,1)$ for $\mu= \cos(\theta)$,}
Finally, for the angular discretization of $\bm{\Omega} = (\theta,\chi) \in [0,\pi] \times [0,2\pi)$, we use a tensor product grid of 16 equally spaced points for the azimuth $\chi$ %(i.e., $\chi_i=(i-0.5)2\pi/16$ for $i=1,\ldots,16$) 
and 8 Gauss-Legendre points for $\mu= \cos(\theta)$, with $\theta$ the inclination.
This yields $4 \times 4$ points per octant and a total of $N_\Omega=128$ discrete directions.
%\cred{\st{In total, for the 4 Stokes parameters, we have $N = 4 N_x N_y N_z N_\Omega N_\lambda \approx 26 \cdot 10^9$ total degrees of freedom.} 
For the four Stokes parameters, the total number of degrees of freedom is $N = 4 N_x N_y N_z N_\Omega N_\lambda = 2.6 \cdot 10^{10}$.

\subsection{Algebraic formulation}\label{sec:algebraic}
Following \citet{janett2021b} and \citet{benedusi2022}, we collect the discretized Stokes parameters and emission coefficients into the vectors $\mathbf{I}\in\mathbb R^N$ and $\pmb{\epsilon}\in\mathbb R^N$, respectively.
We then introduce the linear RT operator $\Lambda \in \mathbb{R}^{N \times N}$, which encodes the solution of Eq.~\eqref{eq:RT}, given an input emissivity $\pmb{\epsilon}$ and a propagation matrix $K$. 
Similarly, the linear scattering operator $\Sigma \in \mathbb{R}^{N \times N}$ encodes the computation of the scattering integral in Eq.~\eqref{eq:emission}, given an input radiation field $\mathbf{I}$. 
The operators $\Lambda$ and $\Sigma$ depend, respectively, on the numerical methods used to solve the RT equation (formal solution) and to approximate the scattering integral (quadrature). 
Equations~\eqref{eq:RT} and \eqref{eq:emission} can then be written in the following matrix form:
\begin{align}
        \mathbf{I} & = \Lambda \, \pmb{\epsilon} + \mathbf{t},
        \label{eq:RT_dis} \\
        \pmb{\epsilon} & = \Sigma \, \mathbf{I} +
        \pmb{\epsilon}^{\text{th}},
        \label{eq:emission_dis}
\end{align}
where $\mathbf{t}\in\mathbb R^N$ represents the radiation transmitted from the boundaries and the vector $\pmb{\epsilon}^{\text{th}}\in\mathbb R^N$ encodes the thermal contribution to emissivity.
%We recall that, 
Substituting Eq.~\eqref{eq:emission_dis} into Eq.~\eqref{eq:RT_dis}, the full non-LTE RT problem can be cast into the linear system:
\begin{equation}\label{eq:linear_system}
        (Id-\Lambda\Sigma)\mathbf{I}=\mathbf{b},
\end{equation}
where $Id$ is the size $N$ identity matrix and $\mathbf{b} = \Lambda \pmb{\epsilon}^{\text{th}} + \mathbf{t}$.

\subsection{Iterative method}
% Solving \eqref{eq:linear_system} with a fixed point iteration, a.k.a. lambda iteration, in the form $\mathbf{I}^{k+1}=\Lambda\Sigma\mathbf{I}^k+\mathbf{b}$, where $k$ is the iteration index, is known to converge slowly (\cite{benedusi2021}).
The iterative solution of Eq.~\eqref{eq:linear_system} is notoriously slow when using a fixed point iteration (or lambda iteration) of the form $\mathbf{I}_{k+1}=\Lambda\Sigma\mathbf{I}_k+\mathbf{b}$, with $k$ the iteration index.
Moreover, standard algebraic preconditioners are not feasible in the considered AD PRD setting due to either their inefficiency or matrix-based nature.
To overcome this limitation, we apply a state-of-the-art Krylov method, namely Flexible GMRES (FGMRES), combined with an innovative physics-based preconditioner obtained by solving the problem in the limit of CRD \citep{benedusi2021,janett2024}.
In particular, we use the parallel implementation of Krylov methods from PETSc\footnote{\url{https://petsc.org/release/}} with a tolerance of $10^{-8}$ for the relative residual.
This iterative method has been shown to clearly outperform standard stationary iterative methods \citep{anusha2009,benedusi2022}.
For a parallel and scalable matrix-free evaluation of $\Sigma$ and $\Lambda$, we employ a distributed strategy based on creating two copies of the discrete field $\mathbf{I}$ with space-based and ray-based domain decompositions, respectively.
This strategy requires a single global communication step between the application of $\Sigma$ and $\Lambda$.
Crucially, the evaluation of each operator is embarrassingly parallel on its corresponding grid.
We remark that for the formal solution, that is the solution of the system of ODEs in \eqref{eq:RT} encoded in \eqref{eq:RT_dis}, we employ a three-point stencil BESSER method for each ODE step \citep{stepan2013}.
%\footnote{The BESSER method includes a conditional statement to avoid overshooting, which can make $\Lambda$ non-linear. In general, a fully linear method should be preferred.}
For the computation of the emissivity \eqref{eq:emission}, encoded in \eqref{eq:emission_dis}, we employ a high-order quadrature and we exploit prior knowledge of the redistribution matrix.
We refer to \cite{benedusi2023} for a comprehensive description of the iterative method, the parallelization strategy, the formal solver, and the quadrature scheme for the scattering integral.

\subsection{Input data} %Input data and parameters

\subsubsection{Atmospheric model}\label{sec:atmosphere}
\begin{figure*}
\centering
	\includegraphics[width=1\textwidth]{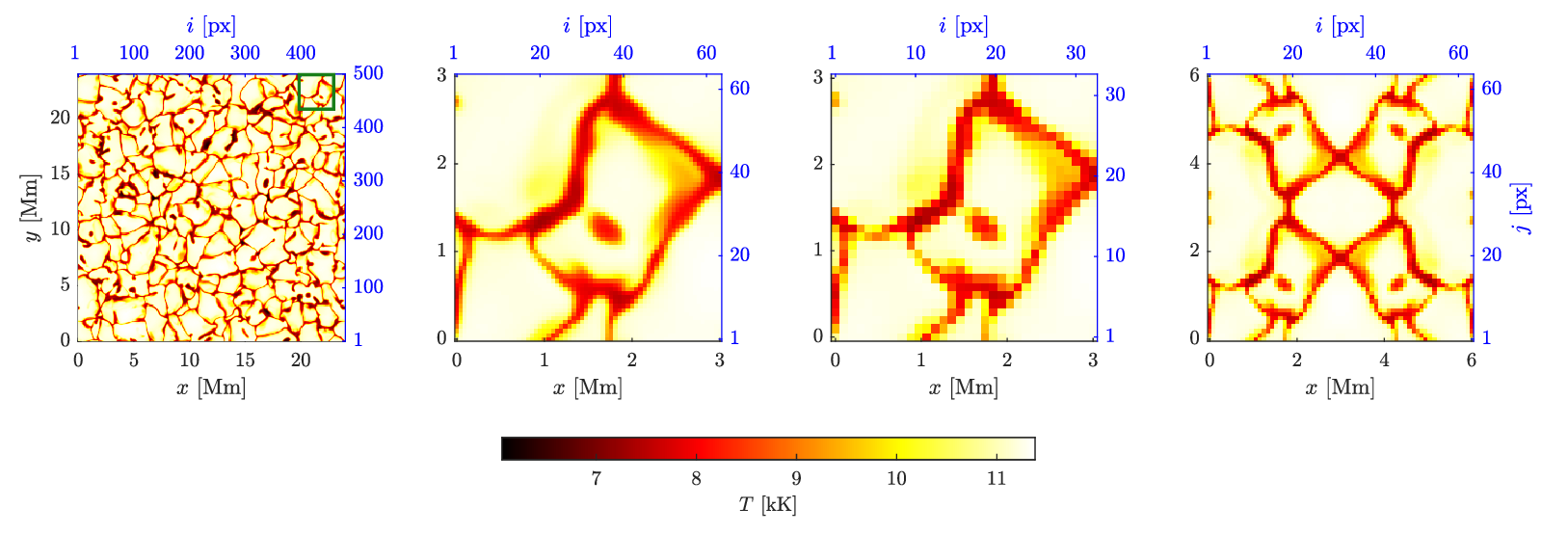}
    \caption{\textit{From left to right:} temperature maps at $z=-0.114$\,Mm
    of (i) the original model from the 3D R-MHD simulation by \citet{carlsson16};
    (ii) a zoomed-in model, corresponding to the green square in the original model;
    (iii) the same zoomed-in model, but with halved spatial resolution in the $x$ and $y$ axes;
    and (iv) Model-63 obtained by mirroring the half-resolution zoomed-in model with respect to the right and top boundaries. 
    }
	\label{fig:atmos_mirror}
\end{figure*}

For this investigation, we use a 3D atmospheric model capturing the structure of the solar photosphere and chromosphere, while being computationally feasible (see Sect.~\ref{sec:discretization}).
To obtain it, we start from a 3D model of the solar atmosphere, including the chromosphere and corona, resulting from the R-MHD simulation of an enhanced network region by \citet{carlsson16}, as shown in the left panel of Fig.~\ref{fig:atmos_mirror}.
This model is based on a spatial grid with $504 \times 504$ nodes on the horizontal planes, covering a region of~$24 \times 24$\,Mm$^2$ with a resolution of~48\,km, and with 496 points on the vertical direction, spanning the height range between $-2.4$\,Mm and 14.4\,Mm. The zero is taken at the average height where the continuum optical depth at 5000\,{\AA} is unity.
Considering the Ca~{\sc i} 4227 formation region, we safely restrict the vertical extension from $-0.114$\,Mm to \mbox{$2.520$\,Mm}, corresponding to 134 grid points with a resolution of 19\,km.
From a quiet region of this model, we extract a smaller box with $64 \times 64$ grid points on the horizontal planes, corresponding to~$3 \times 3$\,Mm$^2$, and we halve its horizontal resolution (see central panels of Fig.~\ref{fig:atmos_mirror}). 
%\cred{\st{The selected region contains a few granules at the photospheric level and is located in a very quiet area, far from the two main magnetic concentrations that characterize the original simulation.}}
To recover periodic boundary conditions, we mirror\footnote{\label{foot:mirror}To guarantee continuous variation of vectorial quantities (i.e., magnetic and bulk velocity fields) across mirror planes, we do not mirror the vectors, but we simply copy their scalar components (magnitude, inclination, and azimuth) at the mirrored spatial points.} the box three-times, obtaining the model shown in the right panel of Fig.~\ref{fig:atmos_mirror}.
% While guaranteeing continuity, this process breaks the mirroring of the vectorial quantities w.r.t. mirror planes.\cred{[@Tanaus\'u check]}}
% To avoid discontinuities, the vectorial physical quantities (i.e., the magnetic field and the plasma bulk velocity) are mirrored keeping... \cred{[Verify with Tanaus\'u how they were mirrored]}.
% }
Noting that the set of grid points at the border between two adjacent boxes are not duplicated, we obtain a periodic 3D model with a spatial grid with $63 \times 63 \times 134$ points, spanning a horizontal region of $6 \times 6$\,Mm$^2$ with a resolution of~96\,km.
This atmospheric model, hereafter referred to as Model-63, is an ideal test-bench for TRIP, allowing us to obtain reliable insights into how the combined effects of PRD and the 3D structure of the solar atmosphere influence the scattering polarization of Ca~{\sc i} 4227.
At each spatial node, the model provides the following physical quantities, needed as inputs for our RT calculations: temperature, number density of electrons and neutral hydrogen atoms, and magnetic field and bulk velocity vectors.

\subsubsection{Atomic model} %\cred{and lower level population [GJ: remove!]}}
We consider a two-level atomic model composed of the upper and lower levels of Ca~{\sc i} 4227.
The upper level (label $u$) has energy $E_u=23652.304$\,cm$^{-1}$, total angular momentum $J_u=1$, and Land\'e factor $g_u=1$, while the lower level (label $\ell$) has $E_\ell=0$\,cm$^{-1}$, $J_\ell=0$, and $g_\ell=0$.
The Einstein coefficient for spontaneous emission is $A_{ul} = 2.18\cdot10^8$\,s$^{-1}$.
% \cred{\st{All values are taken from the online NIST database (REF)} \cluca{[!]}}.

Following the solution strategy described in Sect.~\ref{sec:formulation}, TRIP requires the population of the lower level, that is the ground level of neutral calcium, at each spatial point as a fixed input parameter.
To obtain it, we run the PORTA code in Model-63 for the same two-level atomic model.
We verified, through 1D calculations, that the ground level population of neutral calcium is only marginally affected by PRD effects in Ca~{\sc i} 4227, and it can thus be accurately calculated in the limit of CRD by PORTA,
%This PORTA calculation provides the population of the ground level needed by TRIP, as well as the Stokes profiles of Ca~{\sc i} 4227, in the limit of CRD.   
%We obtain this quantity by solving the non-LTE RT problem for polarized radiation in Model-63, in the limit of CRD, using the PORTA code \citep{stepan2013}.
which in turn needs the number density of neutral calcium atoms as an input parameter.
This is calculated with the HanleRT code \citep{delpino2016,delpino2020}, considering a multi-level atomic model composed of 20 levels for Ca~{\sc i} and the ground level of Ca~{\sc ii}, and solving the non-LTE RT problem column-by-column in Model-63, in the limit of CRD, and neglecting magnetic fields and polarization.
%This calculation is carried out with the HanleRT code \citep{delpino2016}, considering a multi-level atomic model for calcium composed of 20 levels for Ca~{\sc i} and the ground level of Ca~{\sc ii}.
The PORTA and HanleRT calculations are performed considering the problem discretization presented in Sect.~\ref{sec:discretization}.%\footnote{The only exception is the number of azimuths per octant considered by HanleRT, which is one when the model's bulk velocities are neglected (in this case the considered 1D problem has cylindrical symmetry around the vertical) and two when they are included. \cluca{[This footnote can probably be omitted]}}

HanleRT is also used to calculate the broadening constants due to elastic collisions with neutral hydrogen atoms (considering the Van der Waals interaction and quadratic Stark effect) and inelastic collisions with electrons. 
%The number density of neutral hydrogen atoms and electrons is the one provided by Model-63.
HanleRT also provides the depolarizing rate due to elastic collisions \citep[calculated following Sect.~7.13c of][]{LL04}, and the continuum parameters (total opacity, scattering opacity, and thermal emissivity), which are assumed constant with wavelength in the considered spectral interval.

% %
% \begin{figure*}[ht!]
%     \centering
%     \includegraphics[width=0.8\linewidth]{figures/PORTA4}
%     % \includegraphics[width=0.49\linewidth]{figures/PORTA2.png}
%     \caption{Spatial 2D maps of the absolute discrepancy $\widetilde{{\Delta}}_S(i,j)=\max_{\ell,n}(|\Delta_S(i,j,\ell,n)|)$, see Eq.~\eqref{eq:error2}, between the emergent fractional polarization obtained using TRIP and PORTA. 
%     The black star at  $(x,y)=(3.71,1.81)$\,Mm (or $(i,j)=(40,20)$) marks the spatial point discussed in the text.
%     }
%     \label{fig:PORTA4}
% \end{figure*}
% %

\subsection{Computational resources}
\newcommand{\mn}{\footnote{\url{https://www.bsc.es/marenostrum/marenostrum-5}}}
We performed numerical computations with the \textit{MareNostrum 5 GPP}\mn{} (General Purpose Partition) supercomputer, using the \textit{Intel Xeon Platinum 8480+} CPU model. We used one core for each distinct line-of-sight (LOS) and frequency pair, i.e., $N_\Omega N_\lambda=12288$ cores in total, corresponding to 110 nodes. Considering multiple runs with and without magnetic field and bulk velocities, PRD and CRD computations took on average 3h 34m $\pm$ 3m and 36m $\pm$ 5m, respectively.
The storage requirement for a single discrete radiation field
$\mathbf{I}$ in double precision is $8N$ bytes, which corresponds to roughly 210 GB.

\section{TRIP verification}\label{sec:verification}

In this section, we first verify TRIP in the CRD limit, by a comparison with the well-established PORTA code.
Then, we study the impact of varying tolerances and discrete angular grids in PRD. 
To this end, we first introduce a few useful discrepancy metrics.

\subsection{Metrics}
%Let us consider two signals $S$ and $S'$ defined over the 6D computational domain ($\bm{r},\bm{\Omega},\lambda$).
%Given spatial indices $(i,j)\in[1,\ldots,N_x]\times[1,\ldots,N_y]$, a LOS with index $\ell\in[1,\ldots,N_\Omega]$, and a frequency index $n\in [1,\ldots,N_\lambda]$, we define the following 4D absolute metric:\footnote{For notational simplicity we write both $S(x_i,y_j,z_{\max},\bm{\Omega}_\ell,\lambda_n)$ and $S(i,j,\ell,n)$ interchangeably.}
Let us consider two emergent signals $S$ and $S'$ defined over the 5D computational domain with coordinates $(x,y,z=z_{\max},\bm{\Omega}^+,\lambda)$, with $\bm{\Omega}^+$ an outgoing ($\mu>0$) direction.
Given spatial indices $(i,j)\in[1,\ldots,N_x]\times[1,\ldots,N_y]$, a LOS with index $\ell\in[1,\ldots,N_{\Omega^+}]$, with $N_{\Omega^+}=N_{\Omega}/2$ the number of outgoing directions, and a frequency index $n\in [1,\ldots,N_\lambda]$, we define
\begin{equation}\label{eq:error2}
\Delta_S(i,j,\ell,n)=S(i,j,\ell,n)-S'(i,j,\ell,n),
\end{equation}
where, for notational simplicity, we write $S(x_i,y_j,z_{\max},\bm{\Omega}_\ell^+,\lambda_n)$ and $S(i,j,\ell,n)$ interchangeably.
% %%%%
% \cluca{Solo come tentativo:\\
% Let us consider two emergent signals $S$ and $S'$ defined over the 5D computational domain $(x,y,z_{\max},\bm{\Omega}^+,\lambda)$, with $\bm{\Omega}^+$ the outgoing ($\mu>0$) direction.
% Given spatial indices $(i,j)\in[1,\ldots,N_x]\times[1,\ldots,N_y]$, a LOS with index $\ell\in[1,\ldots,N_{\Omega^+}]$, with $N_{\Omega^+}$ the number of outgoing directions, and a frequency index $n\in [1,\ldots,N_\lambda]$, we define the following 4D absolute metric:\footnote{\cluca{For notational simplicity we write both $S(x_i,y_j,z_{\max},\bm{\Omega}_\ell^+,\lambda_n)$ and $S(i,j,\ell,n)$ interchangeably.}}
% %
% \begin{equation}%\label{eq:error2}
% \Delta_S(i,j,\ell,n)=S(x_i,y_j,z_{\max},\bm{\Omega}_\ell^+,\lambda_n)-S'(x_i,y_j,z_{\max},\bm{\Omega}_\ell^+,\lambda_n). \nonumber
% \end{equation}}
% %%%%
For a generic field $X$, we eliminate the frequency dependency, considering the maximum over the frequency domain:
\begin{equation}\label{eq:error3}
\widehat{X}(i,j,\ell)=\max_{n}(|X(i,j,\ell,n)|).
\end{equation}
We finally consider the relative discrepancy, or error-to-signal ratio, for the field $S$:
\begin{equation}\label{eq:error}
\delta_S(i,j,\ell)=\frac{\widehat{\Delta}_S(i,j,\ell)}{\max\{\,\widehat{S}(i,j,\ell),\,\widehat{S}'(i,j,\ell)\,\}},
\end{equation}
i.e. the largest relative difference in $[\lambda_{\min},\lambda_{\max}]$.
% and also the relative discrepancy at peak:
% \begin{equation}\label{eq:error_peak}
% \widehat{\delta}_S(i,j,\ell)=\frac{|\widehat{S}(i,j,\ell) - \widehat{S}'(i,j,\ell) |}{\max\{\,\widehat{S}(i,j,\ell),\,\widehat{S}'(i,j,\ell)\,\}}.
% \end{equation}
% We note that the relative discrepancy at the peak $\widehat{\delta}_S$ is not appropriate for PRD, because of the possible multiple peaks in the PRD profiles, while it is viable in CRD. Moreover, it is not suitable for $S=I$.
%%
We remark that the discrepancy metric defined in Eq.~\eqref{eq:error} becomes uninformative for vanishing signals.

%%%%
% and its maximum in each spatial node:
% \begin{equation}\label{eq:error3}
% \tilde{\Delta}_S(i,j)=\max_{\ell,n}(|\Delta_S(i,j,\ell,n)|).
% \end{equation}

\subsection{Cross-check of TRIP with PORTA}

\begin{figure*}[h]
    \centering
    \includegraphics[width=1\linewidth]{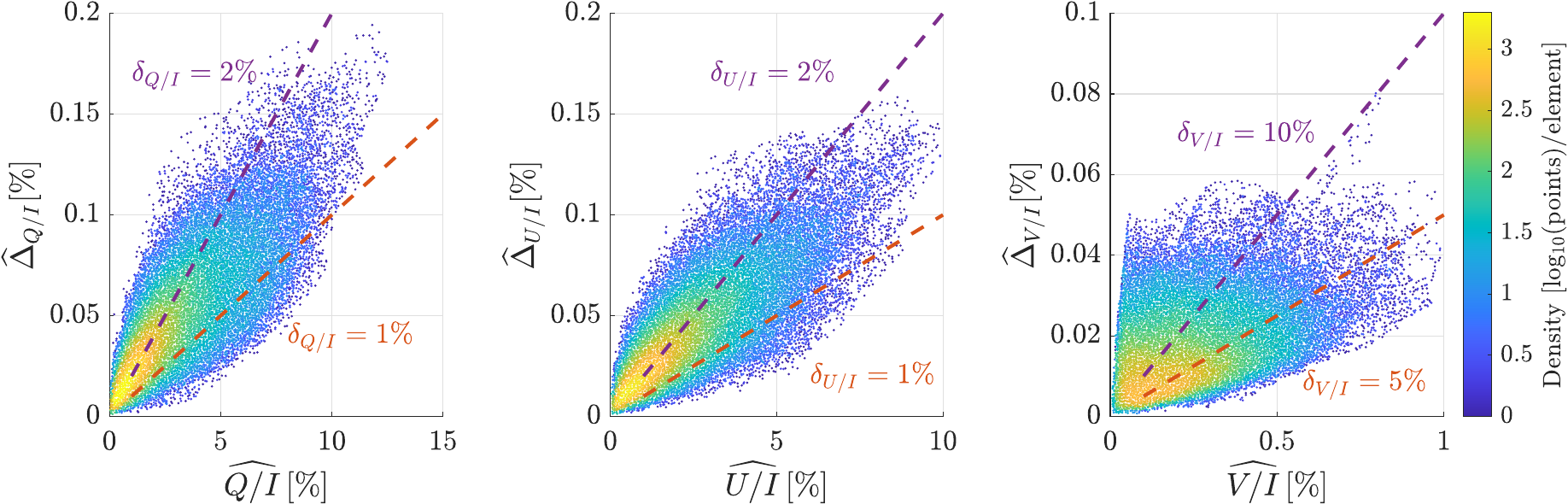}   
    \caption{
    Scatter plots of the absolute discrepancy between TRIP and PORTA emergent fractional polarization as a function  of TRIP fractional polarization, for all $N_x N_y N_{\Omega^+} = 63 \cdot 63 \cdot 64\simeq0.25M$ spatial points and emerging LOSs.
    Dashed lines represent different relative discrepancies $\delta_S$. The colorbar indicates the density of points per grid element, for a $100\times100$ grid of the shown $(\widehat{S},\widehat{\Delta}_S)$ planes.}
    % For each spatial point and LOS in the discrete grid, we plot the maximum polarization signal and the maximum absolute discrepancies. More precisely, for signal $S(i,j,\ell,n)$ we collect all tuples $(|S(i,j,\ell)|,|\Delta_S(i,j,\ell)|)$, according to \eqref{eq:error3}. Slopes corresponding to different signal to error ratios $\delta_S$ are also shown. Colorbars represent density in thousands of points over grid elements, for a $100\times100$ grid of the shown $(|S|,|\Delta_S|)$ planes. For each plot, $N_x\cdot N_y\cdot N_\Omega=63\cdot63\cdot128\approx0.5\mathrm{M}$ points are shown.
    % }
    \label{fig:PORTA5}
\end{figure*}

For verification, we present a comparative analysis between TRIP results in the CRD limit and those obtained from PORTA. Notably, both TRIP and PORTA employ BESSER \citep{stepan2013} as formal solver, use the same discrete grids, and linear interpolation for optical depth conversion. On the other hand,  they differ in the implementation of the quadrature for the scattering integral and in the overall iterative scheme and stopping criteria. 
% In Fig.~\ref{fig:PORTA4}, we show that the absolute differences between the fractional polarization signals obtained with TRIP and PORTA are always bounded by $0.2\%$.
% Averaging Eq.~\eqref{eq:error} over all spatial points and directions, that is over $(i,j,\ell)$, we obtain relative discrepancies of $\bar{\delta}_I=1\%$, $\bar{\delta}_Q=2\%$, $\bar{\delta}_U=2\%$, and $\bar{\delta}_V=5\%$.
% In Fig.~\ref{fig:PORTA2}, we illustrate an example of one of the largest discrepancies found between PORTA and TRIP emerging Stokes profiles, with relative discrepancies $\delta_I=0.7\%$, $\delta_Q=2\%$, $\delta_U=2\%$, and $\delta_V=6\%$. For the fractional polarization, the maximum absolute discrepancies are %$|\Delta_I|=0.007$, 
% $\widehat{\Delta}_{Q/I}=0.15\%$, $\widehat{\Delta}_{U/I}=0.05\%$, and $\widehat{\Delta}_{V/I}=0.03\%$.
% % where large absolute discrepancies $\Delta_{Q/I}$ and $\Delta_{U/I}$ are observed.
% The corresponding spatial point with spatial index $(i,j)=(40,20)$ is labeled with a star in Fig.~\ref{fig:PORTA4}. 
% \cred{\st{In particular, we observe average discrepancies of $\bar{\delta}_I=1\%$, $\bar{\delta}_Q=2\%$, $\bar{\delta}_U=2\%$, and $\bar{\delta}_V=5\%$, averaging Eq.{eq:error} over all spatial points and directions $(i,j,\ell)$.}}
In Fig.~\ref{fig:PORTA5}, we present comprehensive data of absolute discrepancies between emergent fractional polarization signals obtained with PORTA and TRIP and the corresponding TRIP signals. 
We also include slopes with constant relative discrepancy.
Absolute discrepancies are bounded by $0.2\%$ for $Q/I$ and $U/I$, and by $0.08\%$ for $V/I$. 
Moreover, we note that outliers with large relative discrepancies correspond to weak polarization signals. 
In Appendix \ref{appendix}, we present additional details about the TRIP and PORTA cross-check, showing that the absolute discrepancies of signals peaks  are always bounded by $0.07\%$.
In general, we observe a solid agreement between the two codes, noticing that peak relative discrepancies $ \gtrsim 10\%$ are only found for fractional polarization signals $\lesssim0.2\%$.
% especially for fractional polarization signals larger than $0.05\%$.
This comparison simultaneously verifies the accuracy and reliability of both codes in CRD.
Moreover, we note that the differences between the two codes are negligible w.r.t. discretization errors introduced by the angular grid, as shown in the following section.

\subsection{Tolerance}
We now analyze the impact of the relative tolerance on the iterative FGMRES solution of the linear system~\eqref{eq:linear_system}.
% in the scenario with bulk velocities and magnetic fields.
Table~\ref{tab:rtol} reports the effect of varying the tolerance on the maximum relative discrepancies (w.r.t. the solution obtained with a tolerance $\mathrm{tol}=10^{-9}$), the number of FGMRES iterations, and the total runtime.
Notably, a tolerance of $\mathrm{tol}=10^{-7}$ is sufficient to accurately reproduce polarization signals, with errors in $Q/I$, $U/I$, and $V/I$ below $10^{-6}$. % $0.0001\%$.

% \corange{[GJ: old version for comparison.]
% We justify the choice of a relative tolerance $\mathrm{tol}=10^{-8}$ for the linear solution strategy.
% Given \eqref{eq:error}, we compare the emerging surface radiation using multiple tolerances. We consider the most generic atmospheric settings, with nonzero magnetic field and bulk velocities. Considering two discrete solutions obtained using $\mathrm{tol}=10^{-8}$ and $\mathrm{tol}'=10^{-9}$ respectively, we have $\delta_I<10^{-7}, \delta_Q<10^{-4}, \delta_U<10^{-4}$, and $\delta_V<10^{-5}$ for all $i,j=1,\ldots,N_x$ and emerging LOSs. We refer to Table~\ref{tab:rtol} for a more comprehensive comparison.
% For a less accurate solution with tolerance $\mathrm{tol}=10^{-7}$, compared with $\mathrm{tol}'=10^{-9}$, we obtain $\delta_I<10^{-6}, \delta_Q<10^{-3}, \delta_U<10^{-3}$, and $\delta_V<10^{-4}$ for all $i,j=1,\ldots,N_x$.
%
% In Fig.~\ref{fig:rtol}, we show emerging Stokes profiles for a specific spatial point and LOS corresponding to the largest $\delta_Q$.
% With a tolerance of $10^{-7}$, we can capture polarization signals accurately, with errors in $Q/I,U/I$ and $V/I$ that do not exceed $0.0001\%$, compared to an highly accurate solution corresponding to a tolerance of $10^{-9}$.

%
\begin{table*}[h]
    \centering
    \begin{tabular}{c|c|c|c|c|c|c}
         tol & $\max_{i,j,\ell}(\delta_I)$ & $\max_{i,j,\ell}(\delta_Q)$ & $\max_{i,j,\ell}(\delta_U)$ & $\max_{i,j,\ell}(\delta_V)$ & FGMRES its. & runtime\\
         \hline
         $10^{-7}$ & $9.1\cdot10^{-7}$ & $2.1\cdot10^{-4}$ & $6.0 \cdot10^{-4}$ & $7.9\cdot10^{-6}$ & 6 & 3h 24m\\
         $10^{-8}$ & $9.9\cdot10^{-8}$ & $3.5\cdot10^{-5}$ & $4.8 \cdot10^{-5}$ & $1.6\cdot10^{-6}$ & 7 & 3h 39m\\
         $10^{-9}$ & 0 & 0 & 0 & 0 & 8 & 4h 11m
    \end{tabular}
    \caption{
    For different tolerances ($\rm tol$), maximum relative discrepancies w.r.t. the solution obtained with $\rm tol=10^{-9}$, see Eq.~\eqref{eq:error},
    number of FMGRES iterations, and total runtime (using 12288 cores).}
    \label{tab:rtol}
\end{table*}

\begin{table*}[h]
    \centering
    \begin{tabular}{c|c|c|c|c|c|c|c|c}
         $N_\Omega$ & $N$ & $\bar\delta_I$ & $\bar\delta_Q$ & $\bar\delta_U$ & $\bar\delta_V$ & FGMRES its. & runtime & cores\\
         \hline
         72  & $14.7\cdot10^9$ & 0.3\% & 10\% & 20\% & 2\% & 7 & 1h 50m & 6912\\
         128 & $26.1\cdot10^9$ & 0.05\% & 4\% & 5\% & 0.4\% & 7 & 3h 39m & 12288\\
         200 & $40.8\cdot10^9$ & 0 & 0 & 0 & 0 & 7 & 4h 50m & 19200
    \end{tabular}
    \caption{
    For different angular grid sizes $N_\Omega$, degrees of freedom $N$, average relative discrepancies w.r.t. the solution obtained with $N_\Omega = 200$, see Eq.~\eqref{eq:error}, number of FMGRES iterations, total runtime, and number of cores. For all cases,
    there are $2.12\cdot 10^6$ degrees of freedom per core.}
    \label{tab:LOS}
\end{table*}

% \begin{figure}
%     \centering
%     \includegraphics[width=\linewidth]{figures/rtol.png}
%     \caption{First row: emerging Stokes profiles, solutions obtained with $10^{-7}$ and $10^{-9}$ tolerances. Bottom row: absolute errors w.r.t. to an accurate solution in the emerging Stokes profiles, with $\mathrm{Err}(X)=|X_\mathrm{tol}-X'|$, with $X'$ an accurate solution obtained with $10^{-9}$ tolerance. Plots corresponds to the spatial point $(i,j)=(28,25)$ and LOS $(\mu,\chi)=(0.96,6.08)$, where the largest $\delta_Q$ is observed.}
%     \label{fig:rtol}
% \end{figure}

\subsection{Angular discretization}\label{sec:LOS_grid}

We now analyze the impact of the angular grid on the solution of the linear system~\eqref{eq:linear_system}.
% in the scenario with bulk velocities and magnetic fields.
In Table~\ref{tab:LOS}, we show average relative discrepancies, considering angular grids with three, four, and five directions per quadrant, resulting in $N_\Omega=72$, $N_\Omega=128$, and $N_\Omega=200$ respectively.
In particular, we compare solutions for 15 emerging directions of the $N_\Omega=200$ angular grid (namely $(\theta_i,\chi_j)$, with $i=5,\ldots,10$ and $j=1,\ldots,3$).
We additionally report how runtime and number of FGMRES iterations depend on $N_\Omega$, confirming that the solution strategy is robust w.r.t. the angular discretization \citep{benedusi2022}. In terms of convergence, the error decreases when increasing the angular grid size from $N_\Omega = 72$ to $N_\Omega = 128$.

\begin{figure*}[h]
    \centering
    \includegraphics[width=\linewidth]{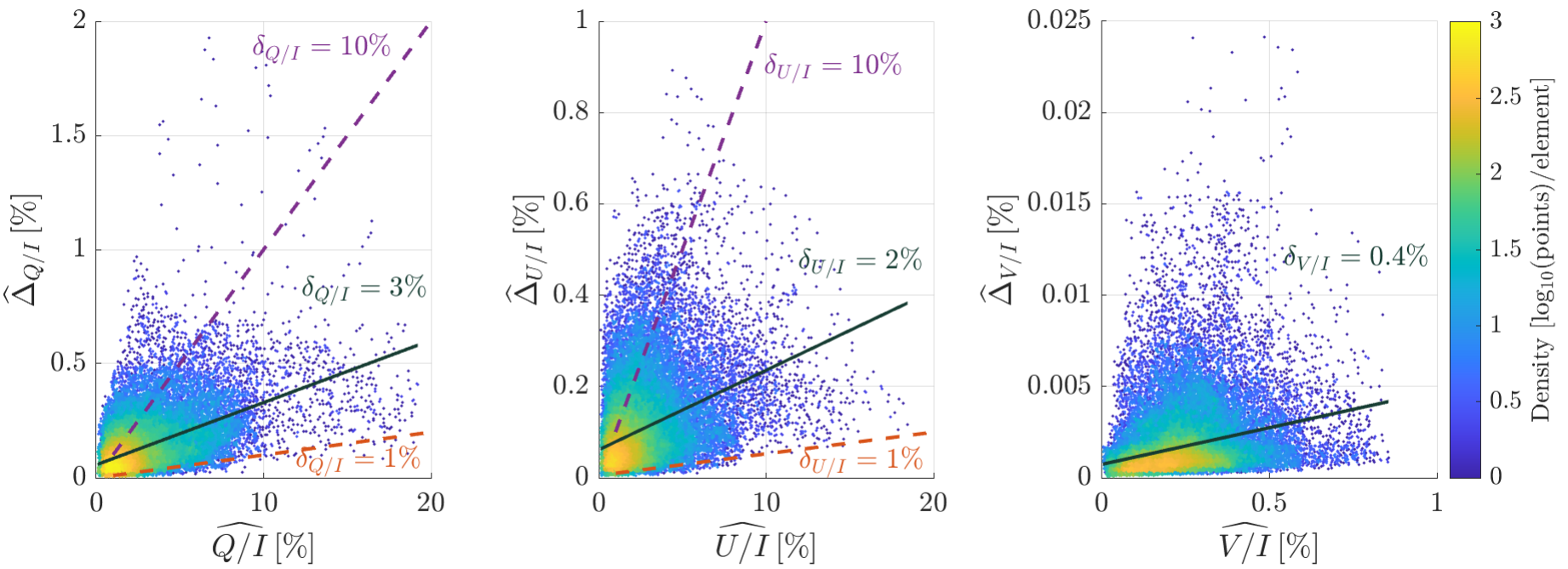}   
    \caption{
    Scatter plots of the absolute discrepancy between the emergent fractional polarization obtained with $N_\Omega=128$ and $N_\Omega=200$ as a function of the fractional polarization for $N_\Omega = 200$. Each plot contains $N_{\widetilde{\Omega}}N_x N_y  = 15\cdot63 \cdot 63 \simeq 60$\,K points, one for each spatial point and common LOS ($N_{\widetilde{\Omega}}=15$ is the number of common LOSs considered for the comparison). 
    Dashed lines represent different relative discrepancies $\delta_S$. Solid lines represent linear regression with corresponding slope $\delta_S$. The colorbar indicates the density of points per grid element, for a $100\times100$ grid of the shown $(\widehat{S},\widehat{\Delta}_S)$ planes.
}
    \label{fig:LOS}
\end{figure*}

\begin{figure}[h]
    \centering
    \includegraphics[width=\linewidth]{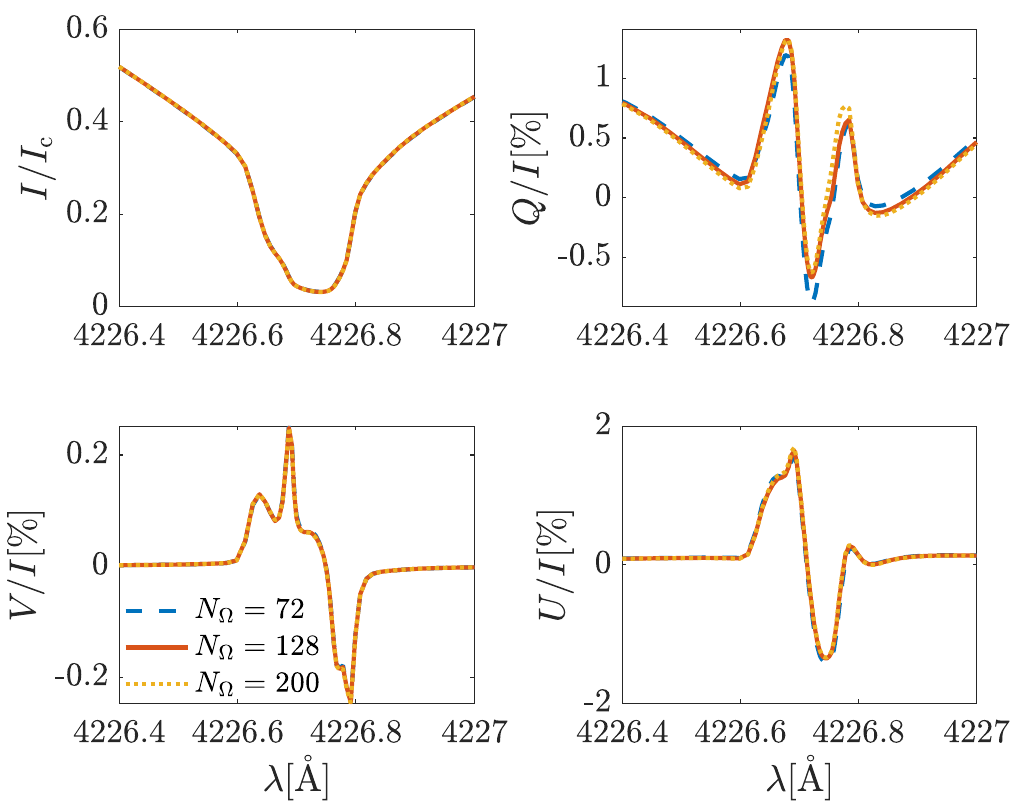} 
    \caption{
    Emergent Stokes profiles at spatial point $(x,y)=(1.43,0.38)$\,Mm with indices $(i,j)=(16,5)$ and LOS $(\mu,\chi)=(0.149,0.157)$, calculated with different angular grid sizes. 
    }
    \label{fig:LOS2}
\end{figure}

In Fig.~\ref{fig:LOS}, we present comprehensive data of absolute discrepancies between fractional polarization signals obtained with $N_\Omega=128$ and $N_\Omega=200$ and the corresponding signals for $N_\Omega = 200$.
To facilitate the identification of an average behavior, we also applied a linear regression model, obtaining relative discrepancies of $\delta_{Q/I}\simeq3\%$, $\delta_{U/I}\simeq2\%$, and $\delta_{V/I}\simeq0.4\%$ (solid lines). 
The measured upper bounds of relative discrepancies are $\delta_{Q/I}=34\%$, $\delta_{U/I}=104\%$, and $\delta_{V/I}=4\%$, all corresponding to small polarization signals.
As an example, in Fig.~\ref{fig:LOS2}, we show profiles corresponding to a LOS and spatial point with $\delta_{Q/I}=20\%$, $\delta_{U/I}=3\%$, and $\delta_{V/I}=1\%$.
% with maximum absolute discrepancies $\widehat{\Delta}_{Q/I}=0.25\%$, $\widehat{\Delta}_{U/I}=0.05\%$, and $\widehat{\Delta}_{V/I}=0.003\%$.
% (w.r.t. $N_\Omega=128$ and $N_\Omega=200$ solutions). 
% \cred{Let us remark that the \corange{\st{maximum difference}} \cluca{[In che senso ``maximum''? Punto spaziale, LOS, e frequenza (posizione del picco), sono fissati, giusto?]} \corange{absolute discrepancy} for $Q/I$ at signal peak is $0.1\%$}.
This figure highlights that large relative errors are essentially located at sharp gradients of relatively small signals\footnote{An in-depth inspection of the results indicates that this behavior is not limited to the present example but represents a general trend.}. 
% corresponding to small oscillating signals.
% \cred{The decrease in error from $N_\Omega = 72$ to $N_\Omega = 128$ clearly indicates convergence with respect to angular grid refinement.}
The angular grid with $N_\Omega=128$ is used consistently throughout this work, as it guarantees an adequate level of accuracy.

%\begin{figure}
%    \centering
    %\includegraphics[width=0.5\linewidth]{figures/weak.png}
    %\caption{Weak scaling increasing $N_\Omega$ of different parts of the solution strategy.}
    %\label{fig:weak}
%\end{figure}

%\FloatBarrier

%\newpage

\section{Results}\label{sec:results}

The radiation synthesized in a 3D model atmosphere contains extensive information encoded in its spatial, angular, and spectral dependencies. 
In this paper, we focus on the radiation emerging along the vertical (i.e., for the LOS with $\mu=1$), corresponding to an observation at the solar disk center.
The reference direction for positive Stokes $Q$ is always taken parallel to the $x$ axis of the chosen coordinate system (see Fig.~\ref{fig:atmos_mirror}).

Scattering polarization signals arise at $\mu=1$ whenever axial symmetry with respect to the vertical is lacking. 
This can be due to (i) deterministic magnetic fields,\footnote{This scenario was first identified in static, plane-parallel 1D atmospheric models, where it has been referred to as forward-scattering Hanle effect \citep{jtb2001hanle}.} (ii) horizontal inhomogeneities in the solar plasma, and (iii) spatial gradients of the plasma bulk velocity \citep[e.g.,][]{jtb2002nat,manso2011,stepan2016}.
%\footnote{We note that only mechanism (i) can be considered in 1D atmospheric models, where it is generally referred to as forward-scattering Hanle effect \citep{jtb2001hanle}.}
% Using the PORTA code, \citet{jaume2021CaI} synthesized the Ca~{\sc i} 4227 disk-center polarization in a 3D model of the solar atmosphere, accounting for all these symmetry-breaking effects.
Using the PORTA code, \citet{jaume2021CaI} synthesized the Ca~{\sc i} 4227 disk-center polarization in the full 3D model of the solar atmosphere shown in the left panel of Fig.~\ref{fig:atmos_mirror}, accounting for all these symmetry-breaking effects.  
However, that work considered the limit of CRD, and was thus focused on line-core signals.
By using TRIP, we extend this analysis to include PRD effects, presenting the first accurate synthesis of both line-core and line-wing Ca~{\sc i} 4227 polarization signals.
We start analyzing the line-core scattering polarization signals, which we also compare with the results of CRD calculations, and then proceed with the wing signals.

\begin{figure*}[ht!]
     \centering
     \includegraphics[width=1\linewidth]{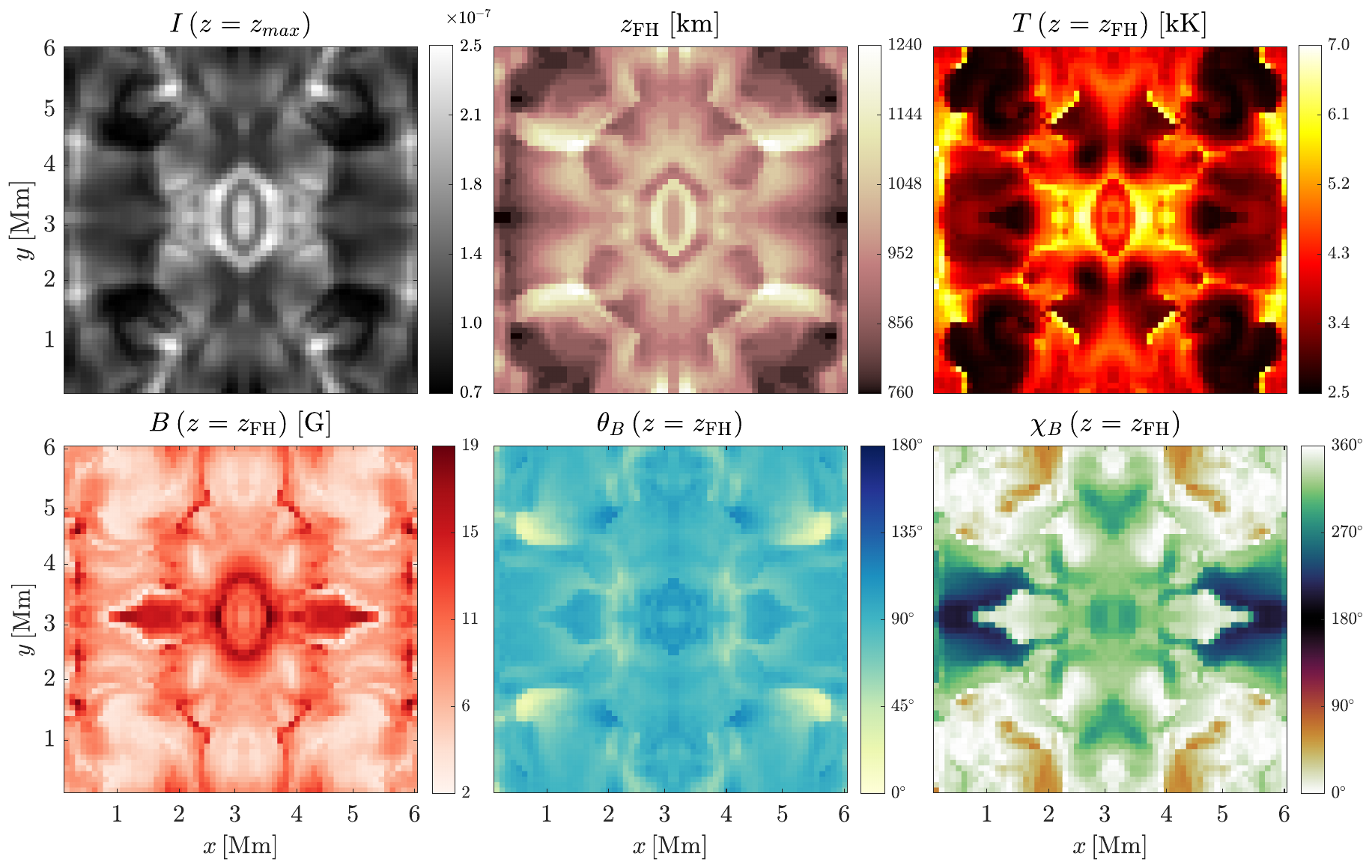}
     \caption{\emph{Upper panels}: Ca~{\sc i} 4227 line-center emergent intensity (in erg cm$^{-2}$ s$^{-1}$ Hz$^{-1}$ sr$^{-1}$) at $\mu=1$ (\emph{left}); line-center formation height (FH) at $\mu=1$ (\emph{middle}); Model-63 temperature at the line-center FH (\emph{right}). 
     % Upper left and middle panels refer to Model-63, 
     Here, $I$ and $z_{\rm FH}$ are calculated
     neglecting magnetic and bulk velocity fields.
     \emph{Lower panels}: Model-63 magnetic field strength (\emph{left}), inclination (\emph{middle}), and azimuth (\emph{right}) at the line-center FH.}
     \label{fig:HeightTepsIc}
\end{figure*}

\subsection{Disk-center core signals}\label{sec:lineCenter}
%
%In the upper left panel of Fig.~\ref{fig:HeightTepsIc}, we first show the intensity of the radiation emergent along the vertical at the Ca~{\sc i} 4227 line-center wavelength, hereafter defined as the wavelength $\lambda_0$ at which the intensity profile has its minimum.
In the upper left panel of Fig.~\ref{fig:HeightTepsIc}, we show the intensity of the Ca~{\sc i} 4227 line-center radiation, emergent along the vertical from Model-63, in the absence of magnetic and bulk velocity fields.
%\cluca{\st{In Model-63, t} T}he line-center formation height, that is the corrugated surface within the 3D model where the optical depth at $\lambda_0$ is unity along the vertical LOS, ranges between about 760\,km and 1240\,km (see upper middle panel of Fig.~\ref{fig:HeightTepsIc}).
The upper middle panel of Fig.~\ref{fig:HeightTepsIc} shows the
corresponding line-center formation height, that is the corrugated surface within the 3D model where the line-center optical depth is unity along the vertical LOS,
which ranges between about 760\,km and 1240\,km.
The line-center wavelength is always defined as the wavelength $\lambda_0$ at which the intensity profile has its minimum.
% In Model-63, neglecting magnetic and bulk velocity fields, it ranges between about 760\,km and 1240\,km.
The comparison between the upper right and left panels of Fig.~\ref{fig:HeightTepsIc} reveals a lack of perfect correlation between the temperature at such heights and the intensity of the emergent line-center radiation, confirming that Ca~{\sc i} 4227 forms under non-LTE conditions.
We also note that at such heights, the magnetic field of Model-63 is relatively weak, with strengths consistently below 20\,G, and with a predominantly horizontal orientation (see lower panels of Fig.~\ref{fig:HeightTepsIc}).

\begin{figure*}[htp!]
    \centering
    \includegraphics[width=0.90\linewidth]{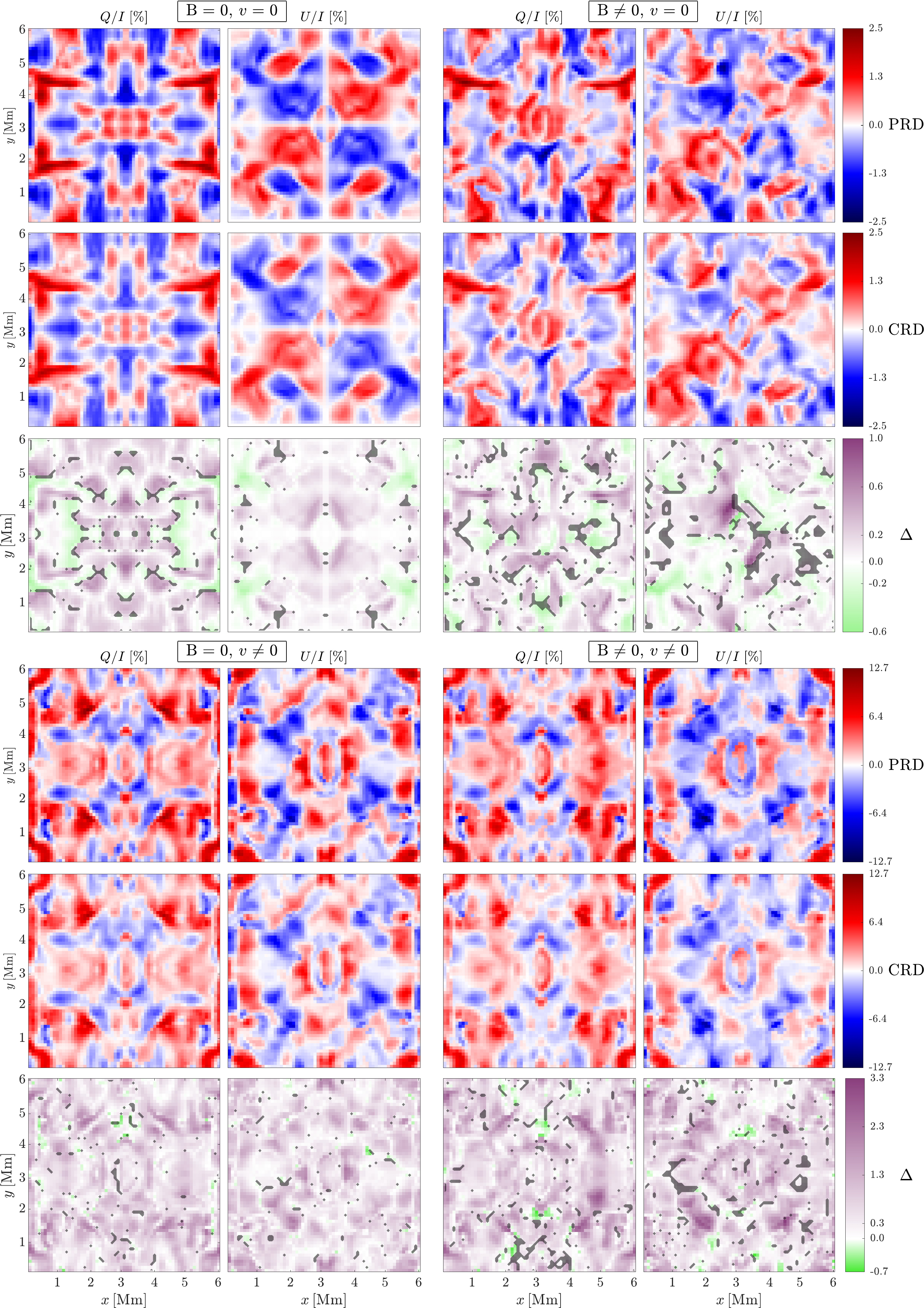}
    \caption{
    Spatial 2D maps for the Ca~{\sc i} 4227 line-core $Q/I$ and $U/I$ signals emergent at $\mu=1$ for both PRD and CRD calculations, obtained by either including or neglecting Model-63 magnetic field $B$ and bulk velocity field $v_B$.
    We also show discrepancy $\Delta$ between PRD and CRD calculations, with the gray areas denoting points with sign discrepancies (NaN in Eq.~\eqref{eq:delta}).
    }
    \label{fig:line-center_maps}
\end{figure*}

% \cluca{\st{We now analyze the scattering polarization signals that PRD effects produce in the line-core of Ca~{\sc i} 4227 at disk center.}}
In Fig.~\ref{fig:line-center_maps}, we present detailed 2D spatial maps of the line-core $Q/I$ and $U/I$ signals emergent at $\mu=1$.
To account for possible shifts and asymmetries in the Stokes profiles induced by the presence of bulk velocities, we extract the maximum absolute value of $Q/I$ and $U/I$ within the spectral interval $\lambda_0 \pm 0.1$\,{\AA}.
Figure~\ref{fig:line-center_maps} is composed of four tiles, each corresponding to a different physical setting, obtained by either including or neglecting Model-63 magnetic and bulk velocity fields.
We note that in the absence of magnetic and bulk velocity fields (see upper left tile), the mirrored structure of Model-63 (see Sect.~\ref{sec:atmosphere}) gives rise to a central symmetry in the polarization pattern across the field of view.
Moreover, it also explains why, taking the reference direction for positive Stokes $Q$ along the $x$ axis, the $U/I$ signals vanish at the domain boundaries and along the median $x$ and $y$ axes.
However, due to the way the vectorial quantities are mirrored (see footnote~\ref{foot:mirror}), the central symmetry of the polarization patterns are partially removed when magnetic and bulk velocity fields are included.

From the top left tile of Fig.~\ref{fig:line-center_maps} (static and unmagnetized case), we observe that the 3D model's horizontal inhomogeneities, which break the axial symmetry of the problem, can generate conspicuous (up to 2.5\%) line-core $Q/I$ and $U/I$ signals with both positive and negative values across the field of view.
%By comparing the left and right tiles, we can see the variation of the $Q/I$ and $U/I$ patterns, due to the rotation of the linear polarization plane produced by the magnetic field via the Hanle effect.
By comparing the left and right tiles, we can appreciate the impact of the magnetic field, via the Hanle effect, on the $Q/I$ and $U/I$ patterns.
A comparison between the upper and lower tiles highlights %the significant increase of the scattering polarization signals (now reaching amplitudes up to 13\%) produced by the spatial gradients of the bulk velocities.
how bulk velocity fields, and in particular their spatial gradients \citep[e.g.,][]{carlin2013}, significantly increase the amplitude of scattering polarization signals (now reaching amplitudes up to 13\,\%).
All these findings are in excellent agreement with the CRD results of \citet{jaume2021CaI}.
The line-core CRD and PRD maps in Fig.~\ref{fig:line-center_maps} actually show a strong qualitative agreement in all considered settings.
For a quantitative comparison, we additionally show in Fig.~\ref{fig:line-center_maps} the discrepancy defined by
\begin{equation}
     \Delta = 
         \begin{cases}
             |S_{\rm \! PRD}| - |S_{\rm \! CRD}|, & \; \text{if} \; \text{sign}(S_{\rm \! PRD}) = \text{sign}(S_{\rm \! CRD}) \\
             \text{NaN}, & \; \text{if} \; \text{sign}(S_{\rm \! PRD}) \ne \text{sign}(S_{\rm \! CRD})\\
         \end{cases}
         \label{eq:delta}
 \end{equation}
with $S=Q/I,U/I$. 
The clear prevalence of violet areas in the $\Delta$ maps shows that CRD calculations generally underestimate the amplitude of the disk-center line-core $Q/I$ and $U/I$ PRD signals in the static case, in agreement with the 1D results of \citet{belluzzi2024}. 
This discrepancy is magnified by the presence of bulk velocities.
Sign discrepancies between PRD and CRD calculations, represented by gray pixels, occur more frequently in the presence of magnetic fields. 
However, these sign discrepancies are typically confined to regions where the polarization is small.
As discussed in \citet{belluzzi2024}, it is not possible to provide a rigorous and comprehensive explanation of the discrepancies between PRD and CRD results.
Noticing that scattering polarization signals depend on the geometry of the scattering and on the detailed angular and spectral properties of the incident radiation field, it seems reasonable that they are lowered by the averages inherent to the CRD approximation, with respect to the general PRD case, in which the detailed coupling between frequencies and propagation directions is accounted for.
On the other hand, it is very hard to make quantitative predictions, especially in nonacademic settings. 
Indeed, the properties of the incident radiation field depend on the thermodynamic structure of the atmospheric model through nonlocal RT effects, and it is extremely difficult both to quantify them and predict how they impact the results under a given approximation.

\subsection{Disk-center wing signals}\label{sec:wing_signal}

\begin{figure*}
    \centering
    \includegraphics[width=\linewidth]{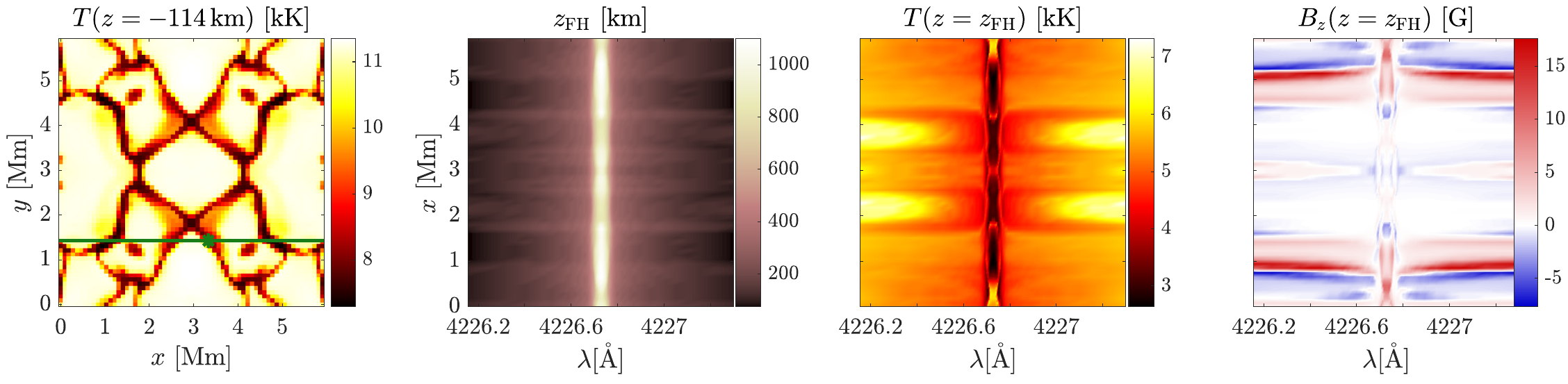}
    \caption{\textit{From left to right}: (i) Model-63 temperature map at $z=-0.114$\,Mm with a green line indicating the row of pixels at $y=1.43$\,Mm (corresponding to index $j=16$) considered for the slit-spectra of Figs.~\ref{fig:slices} and~\ref{fig:slices2}, and a star marking the pixel at $(x,y)=(3.33,1.43)$\,Mm, indexed by $(i,j)=(36,16)$, considered for the profiles of Fig.~\ref{fig:profiles};
    along the considered row of pixels, we show as a function of wavelength (ii) the formation height ($z_{\mathrm{FH}}$) for a vertical LOS, calculated neglecting magnetic fields and bulk velocities; (iii) the temperature at $z_{\mathrm{FH}}$; and (iv) the vertical component of magnetic field at $z_{\mathrm{FH}}$.}
    \label{fig:atm_slit}
\end{figure*}
We now investigate the scattering polarization signals that PRD effects produce in the wings of Ca~{\sc i} 4227, still focusing on the radiation emergent at $\mu=1$.
To this aim, we analyze the Stokes profiles emergent from the row of pixels indicated by the %dashed 
solid green line in the leftmost panel of Fig.~\ref{fig:atm_slit}.
Considering the radiation emergent from such pixels, the second panel from the left of Fig.~\ref{fig:atm_slit} shows that the line formation height drops very quickly when moving from the core to the wings, with the latter forming below 400\,km.
We note that the temperature at the line-core formation height is almost everywhere lower than at the line-wings formation height (see third panel from the left of Fig.~\ref{fig:atm_slit}).
% , suggesting that along the considered \cred{row} of pixels the line core forms close to the model's temperature minimum.
At the formation heights of the wings, the vertical component of the magnetic field $B_z$ (i.e., the longitudinal component for the considered LOS) is very weak, with only a few locations where it reaches values above 15\,G (see rightmost panel of Fig.~\ref{fig:atm_slit}).
We recall that this is the magnetic field component responsible for the magnetic sensitivity of the scattering polarization wings of Ca~{\sc i} 4227 via magneto-optical effects \citep[][]{alsina2018CaI}.

\begin{figure*}[htp!]
    \centering
    \includegraphics[width=\linewidth]{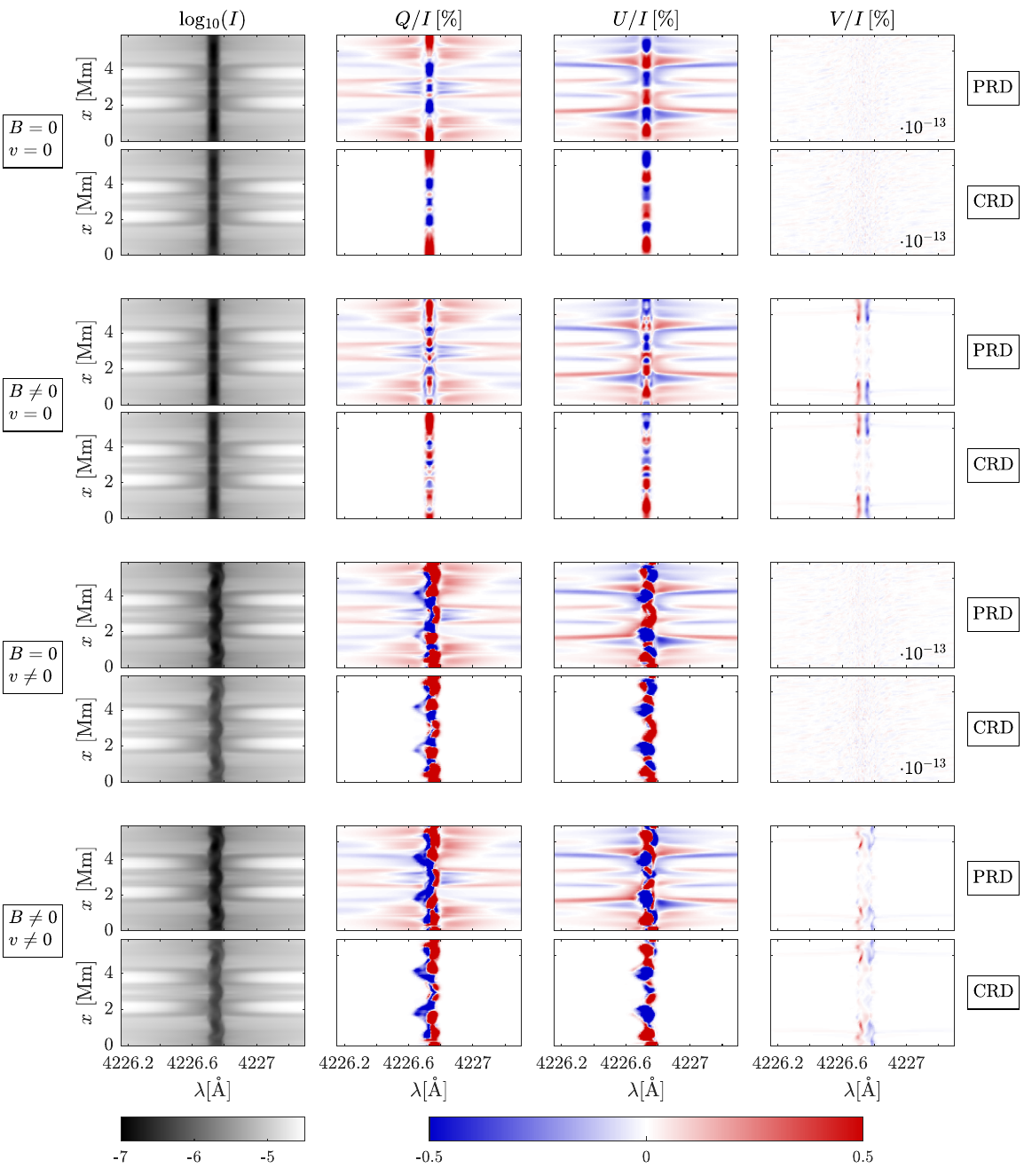}
    \caption{
    % Slit-spectra images (spectral dimension in abscissa and spatial dimension in ordinate) of the four Stokes parameters of the radiation emergent from Model63 at $\mu=1$, in the spectral interval of Ca~{\sc i} 4227, obtained from PRD and CRD calculations, by including or neglecting the model magnetic and bulk velocity fields.
    Slit-spectra for the Ca~{\sc i} 4227 emergent Stokes signals  at $\mu=1$ for PRD and CRD calculations, obtained by either including or neglecting Model-63 magnetic and bulk velocity fields.
    The spatial dimension corresponds to the row of pixels indicated by the green line in the leftmost panel of Fig.~\ref{fig:atm_slit}.
    We note that the color scale is optimized to highlight wing signals, resulting in saturated line-core signals, and that the colorbar has to be intended scaled by $10^{-13}$ for $V/I$ signals with $B=0$, showing the magnitude of the numerical noise. 
    The units for the radiation intensity $I$ are erg\,cm$^{-2}$\,s$^{-1}$\,Hz$^{-1}$\,sr$^{-1}$.
    }
    \label{fig:slices}
\end{figure*}
Figure~\ref{fig:slices} shows slit-spectra of the Stokes parameters emergent from the considered row of pixels at $\mu=1$.
This figure is organized in four tiles, obtained by either including or neglecting the Model-63 magnetic and bulk velocity fields.  
We note that symmetries between the upper and lower parts of the slit-spectra images are a consequence of the mirrored structure of Model-63 (see Sect.~\ref{sec:atmosphere}).
In all tiles, the images for Stokes $I$ clearly show spatial locations with enhanced intensity in the wings, which well correlate with the regions where the temperature at the formation height of the wings is higher (see second panel from the left of Fig.~\ref{fig:atm_slit}).
The upper tile (static and unmagnetized case) shows that the lack of axial symmetry due to horizontal inhomogeneities in the atmospheric model allows PRD effects to produce appreciable wing scattering polarization signals at $\mu=1$.
Such wing signals are instead completely absent in the limit of CRD.
As in the line-core, these signals can be both positive and negative, and have very similar amplitudes in $Q/I$ and $U/I$. 
% 
% A comparison between the two upper tiles (static case) or between the two lower ones (dynamic case) reveals that the magnetic field of Model-63 has no clearly appreciable impact on the wing $Q/I$ and $U/I$ signals.
Comparing the unmagnetized and magnetic cases,
both for the static (two upper tiles) and dynamic (two lower tiles) cases,
we find that the magnetic field of Model-63 has no clearly appreciable impact on the wing $Q/I$ and $U/I$ signals.
This is due to the weak longitudinal component of the magnetic field of Model-63, which is however able to produce appreciable $V/I$ signals in the regions where $B_z$ is strongest.
By comparing the two upper tiles with the two lower ones, we can see that the presence of bulk velocities slightly increases the amplitude of the $Q/I$ and $U/I$ wing signals and also introduces asymmetries.
% when bulk velocities are included.
% As expected, bulk velocities also introduce asymmetries and shifts in the Stokes profiles.
%
%In the core of the line, the $Q/I$ and $U/I$ signals obtained in CRD and PRD can significantly differ as discussed in the previous section.
%This cannot be fully appreciated from Fig.{fig:slices} because the color scale is adjusted to highlight the wing signals, and the core is almost everywhere saturated.
%Nonetheless, where the signal is not saturated (e.g., at the border of the core region), clear differences can be noted between PRD and CRD results.}}
We note that the differences between CRD and PRD calculations in the line-core $Q/I$ and $U/I$ signals, discussed in the previous section, are not fully visible in Fig.~\ref{fig:slices} due to the saturated color scale in the line core.
Finally, no relevant differences between CRD and PRD calculations are present in $V/I$, confirming that
PRD does not significantly affect the circular polarization.

\begin{figure*}
    \centering
    \includegraphics[width=\linewidth]{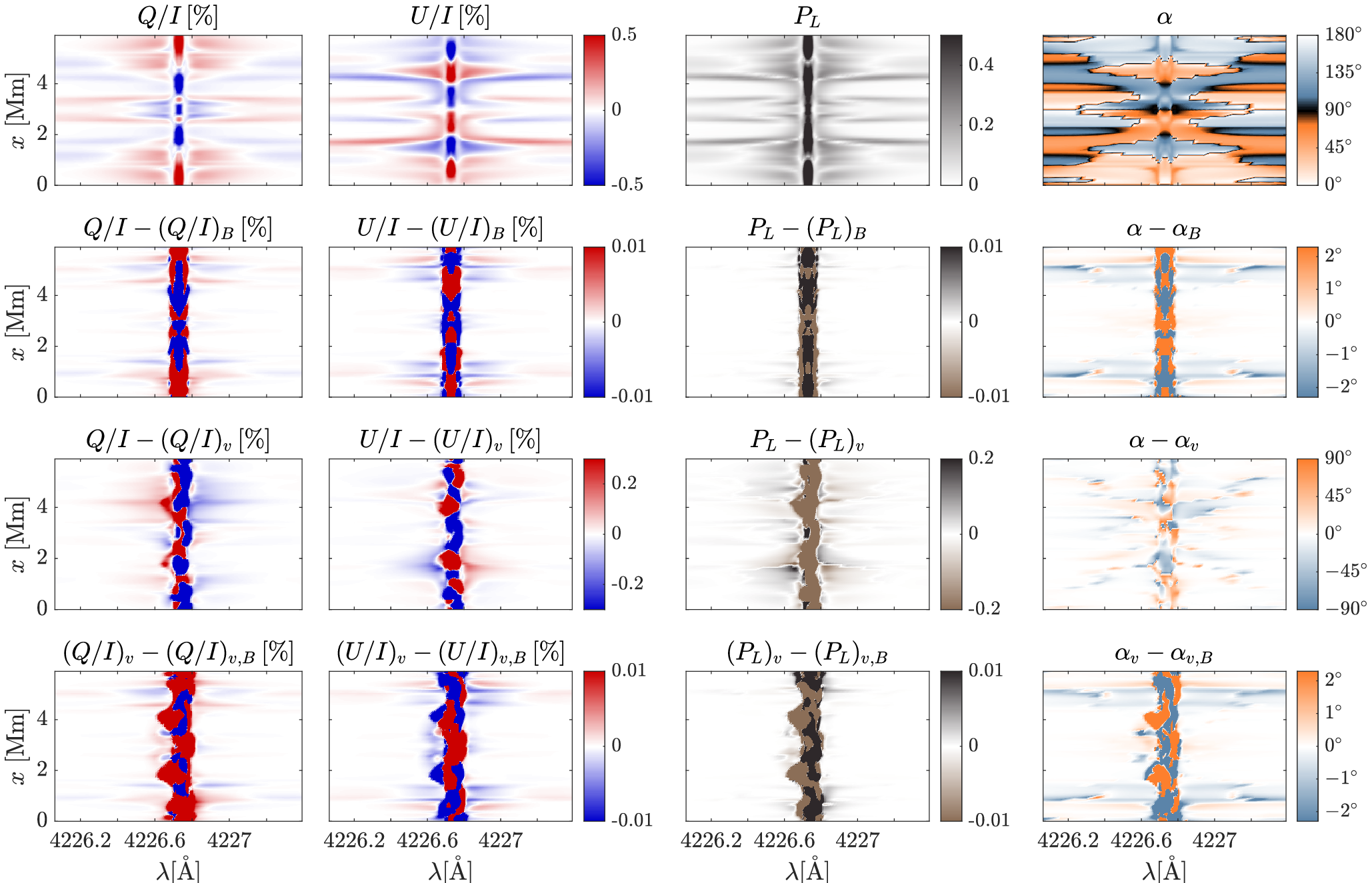}
    \caption{\textit{First row}: slit-spectra for the Ca~{\sc i} 4227 $Q/I$ and $U/I$ emergent signals at $\mu=1$, and corresponding total linear polarization fraction $P_L$ and linear polarization angle $\alpha$ (see Sect.~\ref{sec:wing_signal}), obtained by neglecting Model-63 magnetic and bulk velocity fields.
    %In addition, the corresponding total linear polarization fraction $P_L$ and linear polarization angle $\alpha$ defined in Sect.~\ref{sec:wing_signal}.
    The spatial dimension corresponds to the row of pixels indicated in the leftmost panel of Fig.~\ref{fig:atm_slit}.
    \textit{Second to fourth rows}:
    absolute discrepancies in the same quantities obtained including and neglecting Model-63 magnetic and bulk velocity fields.
    The inclusion of such physical ingredients is indicated through the labels $B$ and $v$, respectively. %indicate that the corresponding quantity is calculated accounting for the magnetic field and the bulk velocity, respectively.}
    We note that the color scale is optimized to highlight wing signals, resulting in saturated line-core signals.
    % absolute difference between the first-row quantities plotted in the first row and the same quantities calculated including Model-63's magnetic field (second row) and bulk velocity (third row).
    % Absolute difference between the same quantities calculated in the presence of bulk velocities, neglecting and including the magnetic field (fourth row).
    % In all panels, the spatial dimension corresponds to the \cred{row} of pixels indicated in panel (a) of Fig.~\ref{fig:atm_slit}; the inclusion of magnetic fields and bulk velocities is indicated by labeling the symbol of a given quantity with the pedices $B$ and $v$, respectively.
    % The color scale is adjusted to highlight the wing signals (core signals are thus generally saturated). 
    }
    \label{fig:slices2}
\end{figure*}
We quantify the effect of magnetic fields and plasma bulk velocities on the line wings by calculating the absolute differences in slit-spectra $Q/I$ and $U/I$ images, including or neglecting such physical ingredients.
Additionally, we extend this analysis to the total linear polarization fraction %given by
%
% \begin{equation}\label{eq:P_L}
$P_L = \sqrt{(Q/I)^2 + (U/I)^2}$,   
% \end{equation}
% 
and to the linear polarization angle $\alpha \in [0^\circ,180^\circ)$ defined as
\begin{equation*}
    \alpha = \frac{1}{2} \tan^{-1} \left(\frac{U}{Q} \right) + \alpha_0 \;\; \text{ with} \; 
    \alpha_0 = 
    \begin{cases}
        0^\circ & \text{if} \; Q > 0 \; \text{and} \; U \ge 0, \\
        180^\circ & \text{if} \; Q > 0 \; \text{and} \; U < 0, \\
        90^\circ & \text{if} \; Q < 0. 
    \end{cases}
\end{equation*}
The second row of Fig.~\ref{fig:slices2} shows that in the line wings the magnetic field, via magneto-optical effects, produces a small rotation ($< 3^\circ$) of the linear polarization plane, as well as a minimal modification of the linear polarization degree.\footnote{We refer to Appendix~A of \citet{alsina2018CaI} for a detailed discussion of the impact of magneto-optical effects on the linear polarization degree.} 
These effects only appear where $B_z$ is strong enough (see rightmost panel of Fig.~\ref{fig:atm_slit}),
% \cluca{right} panel \cluca{\st{d}}),
and induce a slight variation of the amplitude of the $Q/I$ and $U/I$ wing signals (absolute differences below 0.005\%). 
%and only where $B_z$ is strong enough \st{.Here, the magnetic field} to} produce a small depolarization
%\cluca{by reducing the linear polarization degree,} \cred{[GJ: sicuro?] \st{(see grey areas in the line wings in the panel for $P - P_B$)}}, and \cluca{producing} a small rotation ($< 3^\circ$) of the linear polarization plane through magneto-optical effects \cred{\st{(see panel for $\alpha - \alpha_B$)}}.
Interestingly, the direction of rotation perfectly correlates with the sign of $B_z$.
Although the colorbars of line-core signals are saturated, the panel for $P - P_B$ clearly shows that the magnetic field can both polarize or depolarize via the Hanle effect the line-core signals produced by the horizontal inhomogeneities of the model.

The third row of Fig.~\ref{fig:slices2} reveals that bulk velocities appreciably affect the $Q/I$ and $U/I$ line wing-signals (absolute differences up to 0.1\%), although their impact is lower than in the line-core region.
Interestingly, while they generally amplify the line-core linear polarization degree, their impact in the line wings leads to either polarization or depolarization.
Bulk velocities also produce a significant rotation of the linear polarization plane, both in the core (where rotations are generally slightly larger) and in the wings.
%Finally, the last row of Fig.~\ref{fig:slices2} shows that the presence of bulk velocities does not modify the impact of the magnetic field, which remains analogous to the static case, both qualitatively and quantitatively.
Finally, by comparing the second and fourth rows of Fig.~\ref{fig:slices2}, it can be seen that the impact of the magnetic field on the line-wing signals in the static and dynamic cases is qualitatively the same.
\begin{figure*}[h]
\centering
\includegraphics[width=1\linewidth]{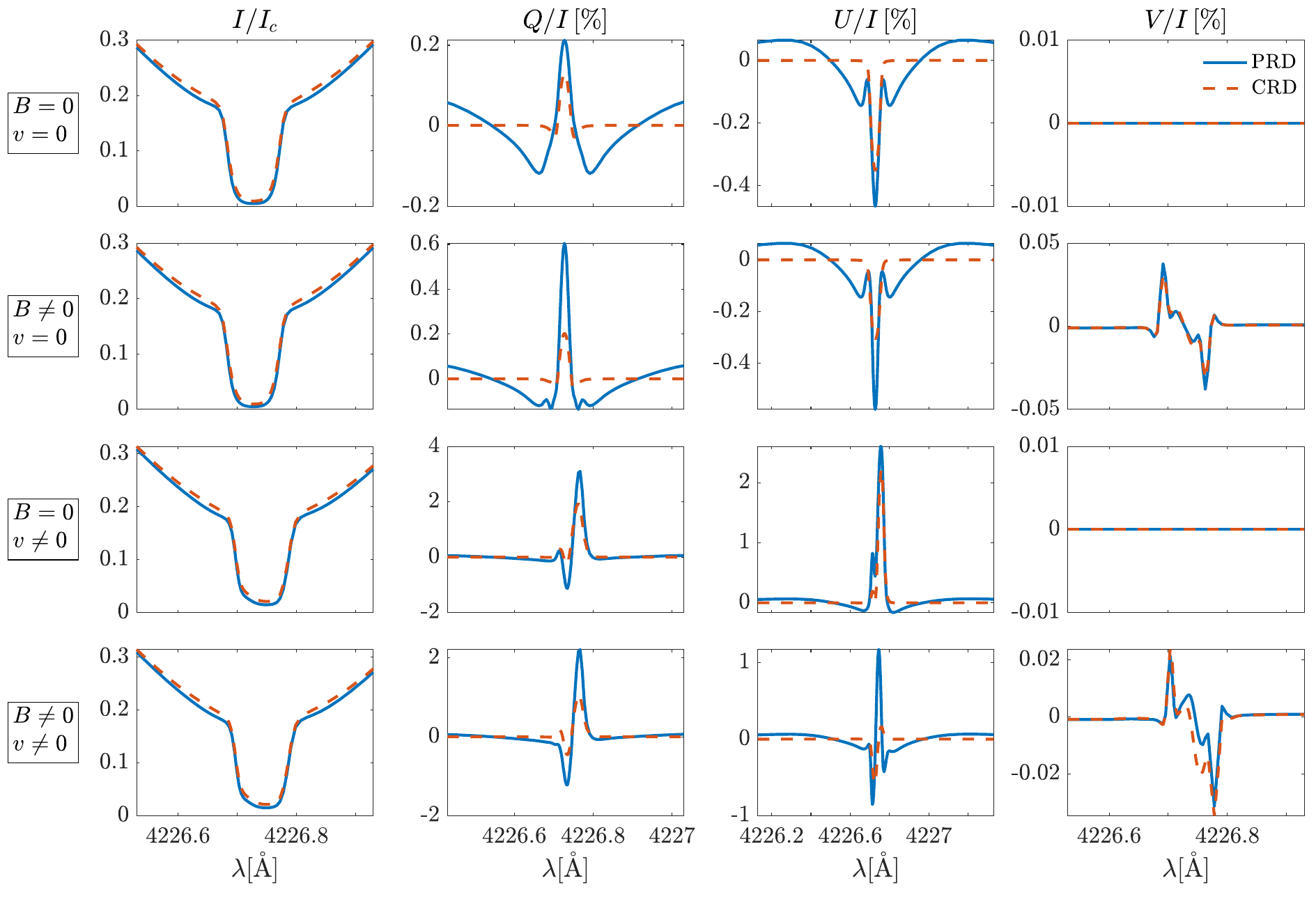}
\caption{
Ca~{\sc i} 4227 Stokes profiles emergent at $\mu=1$ for the spatial pixel indicated by a star in the leftmost panel of Fig.~\ref{fig:atm_slit}.
We compare PRD (solid blue lines) and CRD (orange dashed lines) computations, obtained by either including or neglecting Model-63 magnetic and bulk velocity fields.}
\label{fig:profiles}
\end{figure*}

% \begin{figure*}[]
% \centering
% %\subfloat[][$B=0$ and $v=0$]
% {\includegraphics[width=.85\columnwidth]{figures/profiles_mu1_B0_V0_16_36.pdf}} \qquad \qquad
% %\subfloat[][$B=0$ and $v \ne 0$]
% {\includegraphics[width=.85\columnwidth]{figures/profiles_mu1_B0_V_16_36.pdf}} \\
% %\subfloat[][$B \ne 0$ and $v=0$]
% {\includegraphics[width=.85\columnwidth]{figures/profiles_mu1_B_V0_16_36.pdf}} \qquad \qquad
% %\subfloat[][$B \ne 0$ and $v \ne 0$]
% {\includegraphics[width=.85\columnwidth]{figures/profiles_mu1_B_V_16_36.pdf}}
% \caption{
% Ca~{\sc i} 4227 Stokes profiles emergent at $\mu=1$ for the spatial pixel indicated by a star in the leftmost panel of Fig.~\ref{fig:atm_slit}.
% We compare PRD (solid blue lines) and CRD
% (orange dashed lines) computations, obtained by either including or neglecting Model-63 magnetic and bulk velocity fields. \cred{[spacing]}
% }
% \label{fig:profiles}
% \end{figure*}
%

In Fig.~\ref{fig:profiles}, we show an example of the full Stokes profiles of the radiation emergent at $\mu=1$ from a given point in the field of view (indexed by $(i,j)=(36,16)$ and indicated with a star in leftmost panel of Fig.~\ref{fig:atm_slit}), comparing CRD and PRD calculations, performed either including or neglecting Model-63 magnetic and bulk velocity fields.
The profiles calculated by TRIP in all settings are smooth, with no visible numerical noise or artifacts. 
Moreover, we can identify the signatures of the various physical ingredients, already discussed in the previous sections.
Notably, CRD scattering polarization profiles have lower amplitudes in the line-core
and, for weak signals, can exhibit different shapes and sign with respect to the PRD ones (see, e.g., $U/I$ in the static case).
This emphasizes the crucial role of incorporating PRD when modeling Ca~{\sc i} disk-center polarization signals, both for accurate interpretation and reliable diagnostics.

% Figure~\ref{fig:profiles} shows a comparison between the CRD and PRD Stokes profiles of the radiation emergent at $\mu = 1$ from a given point in the field of view (indicated with a star in leftmost panel of Fig.~\ref{fig:atm_slit}), for different physical settings.
% We can see that the scattering polarization profiles obtained in CRD have generally lower amplitudes than the PRD ones. 
% When the signals are small, they can significantly differ also in shape and sign, with the PRD profiles generally showing more extended wings (see, for instance, the $U/I$ profiles in the static case).
% Interestingly, the depolarization of the line-core $Q/I$ signal produced by the magnetic field, via the Hanle effect, is much stronger in the static case, than in the presence of bulk velocities.
% We also note that the magnetic field reverses the sign of the $U/I$ line-core peak in the static case, and that it can produce (or enhance) secondary peaks next to the central one, in both $Q/I$ and $U/I$.
% Besides increasing the amplitude of scattering polarization signals, the inclusion of bulk velocities (with spatial gradients) generally simplifies the shapes of the $Q/I$ and $U/I$ profiles, by producing a clearly dominant peak.
% We finally note some clear differences between PRD and CRD $V/I$ profiles in the presence of magnetic fields and bulk velocities.

\section{Conclusions}\label{sec:conclusions}

In this work, we presented the first physically relevant results obtained with TRIP, the first code capable of solving the non-LTE radiative transfer problem including scattering polarization and (angle-dependent) PRD effects in comprehensive 3D models of the solar atmosphere.
Based on the theory of \citet{bommier1997b}, TRIP is tailored for resonance lines and allows considering two-level atoms with an unpolarized and infinitely sharp lower level.
As a first application, we run TRIP to model the intensity and polarization of the Ca~{\sc i} line at 4227\,{\AA} in a $63\times63\times134$ 3D model of the solar atmosphere extracted from the R-MHD simulation of \citet{carlsson16}.
We successfully verified TRIP in the CRD limit against the PORTA code, confirming the accuracy of TRIP results while simultaneously providing a cross-verification of both PORTA and TRIP. 
Moreover, we provided quantitative estimates of the errors introduced by the discretization and the iterative scheme. 
From a physical perspective, we extended previous studies, carried out in academic models \citep[e.g.,][]{anusha2010a,anusha2011a,anusha2023}, providing
%we provided %reliable 
new insights into the combined impact that PRD effects and the 3D structure of the solar atmosphere have on the scattering polarization signals of a strong resonance line.
%\cred{So far, this kind of effects had only been explored by \citet{anusha2023}, who modeled an hypothetical spectral line in an academic 3D model of a confined plasma structure.}
In this paper, we focused on the radiation emerging along the vertical, corresponding to an observation at the solar disk center.
A more detailed analysis of the radiation emerging along different lines of sight will be presented in a forthcoming publication.

%Our calculations show that the lack of axial symmetry, due to the horizontal inhomogeneities of the atmospheric model, produces measurable scattering polarization signals at $\mu=1$, both in the line-core region and in the wings.
The most interesting new finding of this work is that when PRD effects are taken into account, the lack of axial symmetry in the problem, due the presence of horizontal inhomogeneities in the atmospheric plasma, can produce measurable scattering polarization signals in the wings of Ca~{\sc i} 4227 at the solar disk center.
% This was already known for the line-core signals, for which our work confirms, and generalizes to PRD, the results of \citet{jaume2021CaI}.
% On the other hand, this is the first time that such disk-center scattering polarization signals are found in the line wings
Our calculations show that these signals can be either enhanced or reduced by bulk velocity gradients and, as expected, they are sensitive to the magnetic field via magneto-optical effects.
% Moreover, we could confirm the magnetic sensitivity of the wing signals via magneto-optical effects.
The signatures of magneto-optical effects are quite small due to the weak strength (and mainly horizontal orientation) of the magnetic field of the considered model, but our 3D results confirm their potential for magnetic field diagnostics.
We additionally note that far-wing signals can be suitably modeled under the lightweight approximation of coherent scattering in the observer frame. This approximation has been recently applied for a 3D RT investigation of the scattering polarization wings of Mg~{\sc ii} h\&k \citep{sukhorukov2025}. %(Sukhorukov et al. in prep.).
Concerning the line-core signals, our calculations confirm and generalize to PRD the CRD results of \citet{jaume2021CaI}. 
We showed that the assumption of CRD generally underestimates the amplitude of these signals, especially in the presence of bulk velocities.
This confirms the findings of \citet{belluzzi2024}, and generalizes them to the 3D and dynamic case.
We confirmed that line-core signals are enhanced by the presence of bulk velocity gradients, and that the magnetic field, via the Hanle effect, modifies them, generally producing a reduction of the linear polarization degree.

The results of this paper demonstrate the capabilities of TRIP in the accurate modeling of scattering polarization in strong resonance lines, representing an important step forward for diagnosing the magnetism of the solar chromosphere and transition region.
% \citep{delacruz_rodriguez2017}. 
Ongoing development of TRIP will make it applicable on larger atmospheric models, allowing a quantitative comparison between simulation results and observational data.
We anticipate extending TRIP to two-term atomic models, thereby broadening its applicability to a wider range of chromospheric lines of significant diagnostic interest, such as Mg~{\sc ii} h \& k, H~{\sc i} Ly-$\alpha$, and the He~{\sc ii} Ly-$\alpha$ line at 304\,{\AA}.

\begin{acknowledgements}
This work was financed by the Swiss National Science Foundation (SNSF) through the Sinergia grant CRSII5-180238.
T.P.A. and J.T.B. acknowledge support from the Agencia Estatal de Investigaci\'on del Ministerio de Ciencia, Innovación y Universidades (MCIU/AEI) under grant
``Polarimetric Inference of Magnetic Fields'' and the European Regional Development Fund (ERDF) with reference PID2022-136563NB-I00/10.13039/501100011033.
T.P.A.'s participation in the publication is part of the Project RYC2021-034006-I, funded by MICIN/AEI/10.13039/501100011033, and the European Union ``NextGenerationEU''/RTRP.  J.\v{S}. acknowledges the financial support from project \mbox{RVO:67985815} of the Astronomical Institute of the Czech Academy of Sciences.
The authors thankfully acknowledge the resources provided by the Barcelona Supercomputing Center in Mare Nostrum 5 GPP through project no. AECT-2024-2-0018. 
This work was also supported by a grant from the Swiss National Supercomputing Centre (CSCS) under project sm74 on Alps and Piz Daint.
\end{acknowledgements}

%Notation for problem discretization
%\begin{align}
%	s & = 1, 2, 3, 4 \qquad \textrm{(Stokes parameters)} \nonumber \\
%	\bm{r}_{\bm{i}} & = (x_i, y_j, z_k) \quad
%	\bm{i}=\bm{1},...,\bm{N} \quad
%	\bm{i}=(i,j,k) \quad
%	\bm{N}=(N_x, N_y, N_z) \nonumber \\
%	\bm{\Omega}_{\bm{l}} & = (\theta_\ell, \chi_m) \quad \quad
%	\bm{l}=\bm{1},...,\bm{M} \quad
%	\bm{l}=(l,m) \quad \;
%	\bm{M}=(N_\theta,N_\chi) \nonumber \\
%	\lambda_n & \qquad \qquad \qquad \; \;
%	n=1,...,N_\lambda \nonumber
%\end{align}

\bibliographystyle{aa}
\bibliography{bibliography}

\appendix
\onecolumn 
\section{TRIP--PORTA comparison}\label{appendix}
% Figure~\ref{fig:PORTA5} provides comprehensive data on the discrepancies between PORTA and TRIP emergent profiles, obtained using the metric defined in Eq.~\eqref{eq:error}. 
% We note two additional remarks\cluca{[discuss]}: (i) the provided data is exhaustive but need further interpretation; (ii) the metric ~\eqref{eq:error} often highlights differences due to minor \textcolor{orange}{\st{profiles} wavelength} shifts, for example due to air-\textcolor{orange}{vacuum} conversions. 
% A complementary metrics, we also consider the relative and absolute discrepancies at the signals peak, i.e. given 
% Eq~\eqref{eq:error3},

\begin{figure*}[h]
    \centering
    \includegraphics[width=0.95\linewidth]{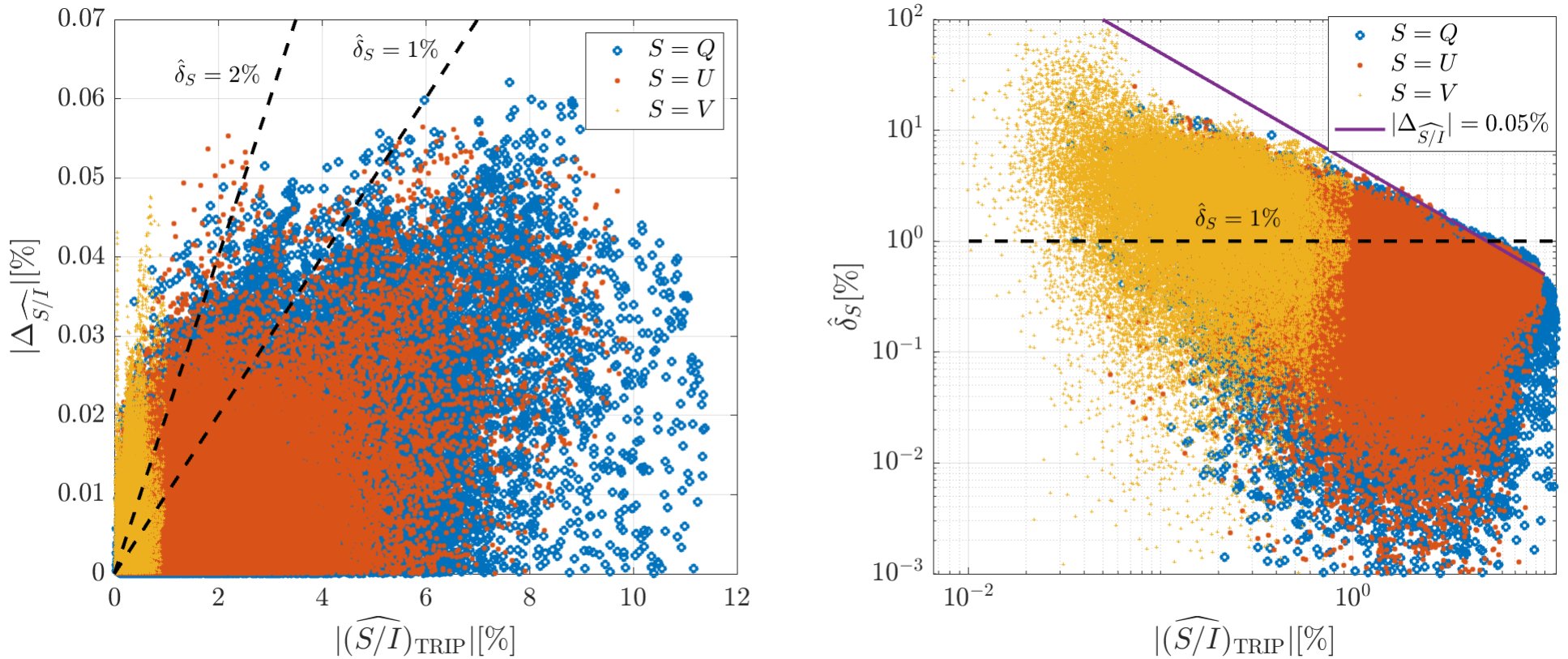}
    \caption{Discrepancies between PORTA and TRIP at peak. Left: absolute discrepancy at peak versus corresponding TRIP peak signals. 
    Slopes corresponding to $\hat\delta_S=1\%$ and $\hat\delta_S=2\%$ are shown. 
    Right: log--log relative discrepancies at peak versus corresponding TRIP peak signals. 
    We show the $\hat\delta_S=1\%$ dashed line as well as a solid line corresponding to a systematic discrepancy of $0.05\%$.}
    \label{fig:appendix1}
\end{figure*}

\begin{figure*}
    \centering
    \includegraphics[width=\linewidth]{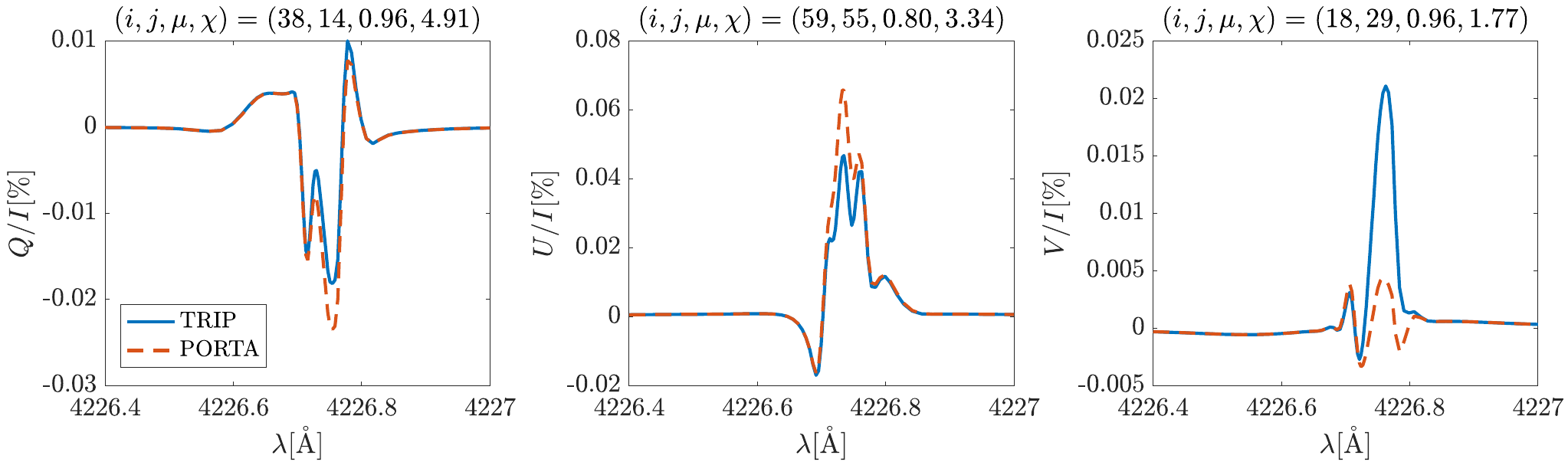}
    \caption{Emergent fractional polarization profiles with the largest relative peak discrepancies, namely, $\hat\delta_Q=30\%$, $\hat\delta_U=41\%$, and $\hat\delta_V=80\%$. Corresponding spatial points and LOSs are reported.}
    \label{fig:appendix2}
\end{figure*}

\begin{figure*}
    \centering
    \includegraphics[width=\linewidth]{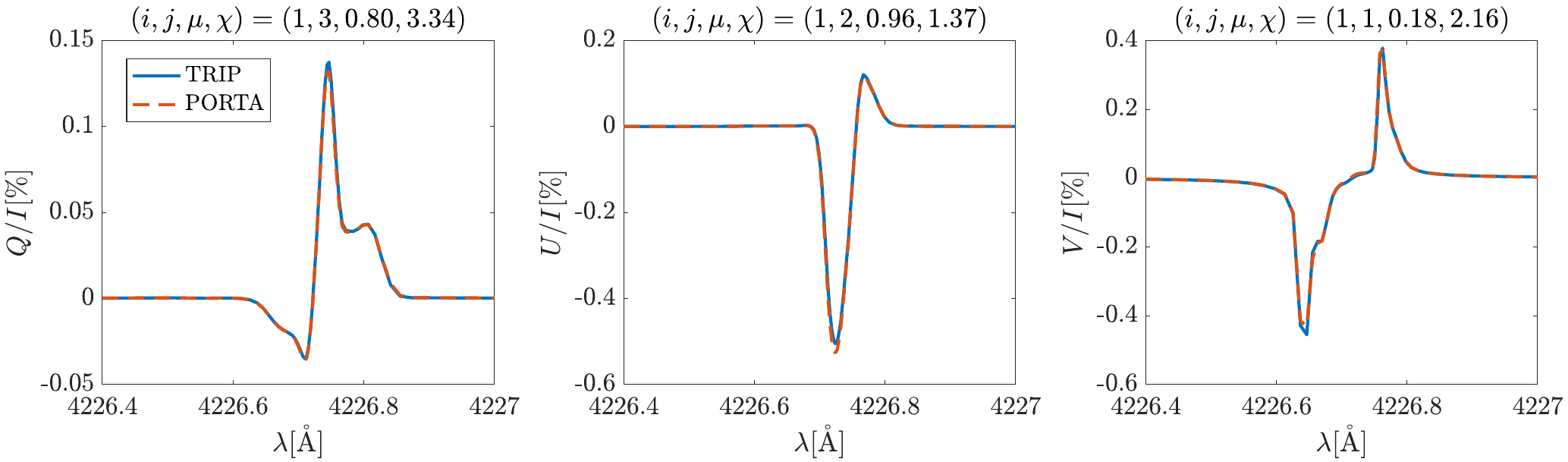}    %\includegraphics[width=0.32\linewidth]{figures/dU_4_percent.eps}
    \caption{Examples of emergent fractional polarization profiles with ``large'' relative discrepancies, namely, $\hat\delta_{Q}=\hat\delta_U=4\%$ and $\hat\delta_V=5\%$. Corresponding spatial points and LOSs are reported.}
    \label{fig:appendix3}
\end{figure*}

In this appendix, we further discuss and interpret the discrepancies between PORTA and TRIP emergent Stokes profiles.
Figure~\ref{fig:PORTA5} provides exhaustive data on such discrepancies, following the metric defined in Eq.~\eqref{eq:error}.
Although such metric is very general, we note that it is prone to emphasize differences due to minimal (less than 1\,m{\AA}) wavelength shifts, 
%for example due to air-vacuum conversions.
of numerical origin, between the TRIP and PORTA profiles.
To avoid this issue, we introduce here a complementary metric that considers the relative and absolute discrepancies at the signals peak.
Given Eq.~\eqref{eq:error3}, this new metric is defined as
\begin{equation*}
\Delta_{\widehat{S/I}} = (\widehat{S/I})_\mathrm{TRIP}-(\widehat{S/I})_\mathrm{PORTA},\quad \hat{\delta}_S=\frac{|\Delta_{\widehat{S/I}}|}{|(\widehat{S/I})_\mathrm{TRIP}|}\quad \mathrm{for}\quad S\in\{Q,U,V\}.
\end{equation*}
This metric is experimentally relevant and can be safely used in the CRD limit, as in this case the Stokes $Q/I$ and $U/I$ profiles have in general one dominant peak.
In Figure~\ref{fig:appendix1}, we show relative and absolute peak discrepancies. 
From the left panel, we see that absolute discrepancies are bounded by $0.07\%$ for all Stokes parameters. 
Moreover, the vast majority ($\approx83\%$) of $Q/I$ and $U/I$ points have a relative discrepancy below $1\%$.
Larger relative discrepancy can be found especially for signals with small amplitudes (e.g. $V/I$ ones). 
%Larger relative discrepancies are only found in signals with small amplitudes.
We clearly show this dependency in the right panel of Figure~\ref{fig:appendix1}, where a systematic discrepancy of $0.05\%$ is also represented as a solid line. 
In Figure~\ref{fig:appendix2}, we show the profiles corresponding to the largest relative discrepancies, highlighting that these appear for small fractional polarization signals, with $\widehat{S/I}\leq 0.06\%$.
Finally, in Figure~\ref{fig:appendix3}, we show emergent profiles corresponding to ``large'' relative discrepancies, i.e. $\hat\delta_Q=\hat\delta_U=4\%$ (larger than 99.4\% of $U/I$ and $Q/I$ profiles) and $\hat\delta_V=5\%$ (larger than 87\% of $V/I$ profiles). We note that a relative peak discrepancy up to $5\%$ corresponds to profiles that are identical for any practical purpose.
As a reference, average relative discrepancies (over all the emergent LOSs and spatial points) are $\hat\delta_Q=\hat\delta_U=0.6\%$ and $\hat\delta_V=2.9\%$. Finally, we remark that discrepancies are further reduced if the spatial resolution is degraded by averaging the profiles over the whole field of view. 
%signals are averaged in space, accounting for the limited resolution of observations. 
In this case, we obtain the following bounds on relative discrepancies $\hat\delta_V<0.02\%$ and $\hat\delta_Q,\hat\delta_U<0.005\%$.
% In conclusion, since PORTA and TRIP are independent and use different numerical schemes, the agreement between their outputs is acceptable.
%

\end{document}